\documentclass[preprint,12pt,3p]{elsarticle}
\journal{Ultramicroscopy}

\usepackage{amsfonts}
\usepackage{amsmath}
\usepackage{gensymb}
\usepackage{amssymb}
\usepackage{siunitx} 
\usepackage[normalem]{ulem}
\usepackage{caption}
\usepackage{subcaption}
\usepackage{dcolumn}
\usepackage{bm}
\usepackage{color}
\usepackage{hyperref}
\usepackage{bookmark}
\usepackage{empheq}
\usepackage{booktabs}
\usepackage{multirow}
\usepackage{pdfpages}
\usepackage{graphicx}
\usepackage{tabu}
\usepackage[version=4]{mhchem}
\usepackage{soul}
\usepackage[utf8]{inputenc}
\usepackage{epigraph}
\usepackage{physics}
\usepackage{textcomp}
\usepackage{tikz}
\usepackage{notoccite}
\usetikzlibrary{shapes.geometric,arrows}
\usepackage[T1]{fontenc}
\usepackage{booktabs}
\usepackage{mathptmx}     
\usepackage{latexsym}
\usepackage{xcolor} 
\usepackage{colortbl}
\usepackage{amsmath}
\usepackage {mathrsfs}
\newcommand{\angstrom}{\textup{~\AA}}

\usepackage[running]{lineno}

\begin{document}

\begin{frontmatter}

\title{Relativistic EELS scattering cross-sections for microanalysis based on Dirac solutions}

\author[EMAT,Nano,Oxford]{Zezhong Zhang}\corref{cor1}
\ead{zezhong.zhang@uantwerpen.be}
\author[RFI]{Ivan Lobato}
\author[Bio21]{Hamish Brown}
\author[EMAT,Nano]{Dirk Lamoen}
\author[EMAT,Nano]{Daen Jannis}
\author[EMAT,Nano]{Johan Verbeeck}
\author[EMAT,Nano]{Sandra Van Aert}\corref{cor2}
\ead{sandra.vanaert@uantwerpen.be}
\author[Oxford]{Peter D. Nellist}\corref{cor3}
\ead{peter.nellist@materials.ox.ac.uk}

\address[EMAT]{Electron Microscopy for Materials Research (EMAT), University of Antwerp, Groenenborgerlaan 171, 2020 Antwerp, Belgium}
\address[Nano]{NANOlab Center of Excellence, University of Antwerp, Groenenborgerlaan 171, 2020 Antwerp, Belgium}
\address[Oxford]{Department of Materials, University of Oxford, 16 Parks Road, Oxford OX1 3PH, United Kingdom}
\address[RFI]{Rosalind Franklin Institute, Harwell Research Campus, Oxfordshire OX11 0FA, United Kingdom}
\address[Bio21]{Ian Holmes Imaging Centre, Bio21 Molecular Science and Biotechnology Institute, University of Melbourne, Victoria, Australia}
\cortext[cor1]{Corresponding author}

\begin{abstract}
The rich information of electron energy-loss spectroscopy (EELS) comes from the complex inelastic scattering process whereby fast electrons transfer energy and momentum to atoms, exciting bound electrons from their ground states to higher unoccupied states. To quantify EELS, the common practice is to compare the cross-sections integrated within an energy window or fit the observed spectrum with theoretical differential cross-sections calculated from a generalized oscillator strength (GOS) database with experimental parameters. The previous Hartree-Fock-based and DFT-based GOS are calculated from Schrödinger's solution of atomic orbitals, which does not include the full relativistic effects. Here, we attempt to go beyond the limitations of the Schrödinger solution in the GOS tabulation by including the full relativistic effects using the Dirac equation within the local density approximation, which is particularly important for core-shell electrons of heavy elements with strong spin-orbit coupling. This has been done for all elements in the periodic table (up to Z = 118) for all possible excitation edges using modern computing capabilities and parallelization algorithms. The relativistic effects of fast incoming electrons were included to calculate cross-sections that are specific to the acceleration voltage.  We make these tabulated GOS available under an open-source license to the benefit of both academic users and to allow integration into commercial solutions.
\end{abstract}


\end{frontmatter}


\section{Introduction}

Electron energy-loss spectroscopy (EELS) involves analyzing the energy loss distribution of electrons after their interaction with a specimen. With the underlying inelastic electron scattering event, the fast-moving incident electron transfers part of its energy and momentum by exciting the atomic system from its initial quantum ground state to various excited final states \cite{egerton2011electron}. Since these transitions are characteristic of the type of elements and their local electronic structure, EELS can fingerprint the composition and bonding of the specimen. Thanks to advances in instrumentation such as aberration correction \cite{rose2009historical}, monochromation \cite{krivanek2014towards,krivanek2014vibrational} and more recently direct electron cameras \cite{hartDirectDetectionElectron2017a,plotkin2020hybrid}, EELS can now be performed with high energy resolution (up to meV) and spatial resolution (up to \AA). Recognized as a powerful analytical tool for site-specific materials characterization \cite{Muller2008,hachtel2019identification}, there is a general interest in the accurate and precise quantification of EELS and this therefore necessitates a reliable theoretical model to calculate the inelastic scattering cross-sections, which describes the probability of inelastic scattering events to take place.

To quantify EELS, one has to model the spectra as a superposition of the background signal and multiple excitation edges with characteristic shapes. The traditional method involves background removal using a power-law extrapolation from the pre-edge region and integration of the edge signal within a certain energy window \cite{egerton2002improved}. For the analysis of fine structures, deconvolution with the low-loss spectrum is performed to remove plural scattering (known as the Fourier-ratio method)\cite{egerton2011electron}. However, this methodology can result in various artifacts during deconvolution and can be very sensitive to the choice of suitable windows for background removal and signal integration \cite{verbeeck2004model}. A better approach is to perform a model-based fitting of the observed spectrum with theoretical scattering cross-sections. This has clear benefits: (a) effective separation of overlapping components, (b) consideration of multiple scattering through convolution of the core-loss model with the low-loss spectra instead of deconvolution, and thereby its artifacts, (c) most significantly, the model-based parameter estimation theory grants statistical superiority, resulting in improved accuracy and free from bias  \cite{leapman1988separation, verbeeck2004model}. The model-based method is implemented in EELSModel \cite{verbeeck2004model} (and its python version pyEELSModel \cite{daen_jannis_2024_10992986}), HyperSpy \cite{de2017electron} (the EDX/EELS analysis is recently renamed exspy) and Gatan Microscopy Suite Software$^{TM}$, which are widely used in the EELS community. Continuous efforts have been devoted to improving the model-based EELS, including developing a better background model \cite{cueva2012data,van2023convexity}, addressing the correlated noise effects from the camera \cite{Verbeeck2008} and incorporating a phenomenological model for the fine structure during signal processing \cite{Verbeeck2006}. However, the day-to-day scattering cross-section computation has long relied on the Gatan generalized oscillator strength (GOS) database, which involves some limitations: (1) the tabulation does not distinguish spin up/down, which can be critical for heavy elements with strong spin-orbit coupling (SOC); (2) not all (though a good coverage) elements and edges are included in the database; (3) the energy-momentum space is limited and sampling is not fine enough for accurate cross-section calculations; (4) the database is under a commercial license limiting its use in community-driven development.

Theoretical core-shell cross-sections are computed based on a set of experimental parameters (such as acceleration voltage, convergence angle, EELS collection angle, etc) and a GOS database, which couples the atomic physics with electron scattering. The name "generalized oscillator strength" comes from the extension of the "oscillator strength", which describes the probability of absorption or emission of electromagnetic radiation caused by the electron orbital transitions for an atom or molecule \cite{schattschneider1986fundamentals}. Indeed the precise analysis of atomic light spectra, tracing back to the starting point of quantum mechanics, lays the foundation of atomic physics. The transition probability can be calculated using Fermi's golden rule, which states that the transition probability is proportional to the strength of the coupling between the initial and final states under weak perturbation (first Born approximation). Interestingly, the formulation was first proposed by Dirac in 1927 \cite{dirac1927quantum}, just a year before the Dirac equation \cite{dirac1928quantum}. On its importance for quantum transitions, Fermi referred to it as the "golden rule No. 2" in a university nuclear physics class \cite{fermi1950nuclear} and thereafter was known as the "Fermi's golden rule". In the context of core-shell electrons, the initial state represents the occupied bound state of a given orbital, while the final states can include unoccupied bound orbitals above the Fermi level or continuum states above the vacuum level after ionization. The local density of unoccupied states highly depends on the valence electron redistribution in the crystal field (known as bonding), leading to complicated energy loss near-edge fine structures (ELNES). In addition, the reflection of outgoing excited electron waves by surrounding atoms will generate the extended energy loss fine structures (EXELFS). In contrast, at a moderate higher energy loss (typically tens of eV above the edge to be free from ELNES and EXELFS), the perturbation caused by the local environment is much decayed, and the excitation to the continuum states defines the general spectra shape for each edge \cite{egerton2011electron}. Thus, the GOS can be computed for a single isolated atom with final states limited to the continuum state and applied to any material system for the quantification of such an element. 

The computation of the theoretical EELS cross-sections dates back to 1932 when Bethe first used the Born approximation to consider the angular-dependent probability of a collision of an electron with matter at relativistic speed using hydrogenic solutions\cite{bethe1932bremsformel}. In the 1950-60s, ionization cross-section studies still often relied on the hydrogenic solutions \cite{moiseiwitsch1968electron}, which describe the intermediate and large energy loss fairly well for K-shells but cannot capture the near-edge behaviours for higher shells. At the same time, more realistic atomic electron wavefunction calculations became increasingly accessible due to the advancements in computing power and more efficient ways of describing the many-body electron-electron interactions. In particular, the exchange contribution which was accounted exactly through the Slater determinant\cite{slater1929} in the Hartree-Fock approach, but was at the same time computationally expensive, was approximated by a density-dependent functional \cite{slater1951}. This charge density approach \cite{thomas1927calculation, dirac1930note, slater1951} has been incorporated into various atomic structure codes, including those developed by Hermann and Skillman \cite{herman1963atomic}, Liberman \cite{liberman1971}, Grant \cite{grant1970}, and Cowan \cite{cowan1981theory}, which were widely used within the spectroscopy community. Over time, this approach evolved into the more generalized framework of density functional theory (DFT) for many-atom systems \cite{kohn1965}. With realistic wavefunctions available, since the 1970s, improvements were made by McGuire \cite{mcguire1971inelastic}, Manson \cite{Manson1972} and Scofield \cite{scofield1978k}, who used Hartree-Fock (HF) solutions for the atomic orbitals to calculate K and L-edge cross-sections of light elements. These HF-GOS efforts were systematically extended to a much broader range of elements and shells by Leapman, Rez and co-workers \cite{Leapman1980a, Saldin1987, Rez1989}. In the early 2000s, Rez \cite{Rez2002,Rez2003} further incorporated the Dirac-Slater program of Liberman\cite{liberman1971} for the core-shell wave function computation, which became the basis for the Gatan GOS database. Many studies in the atomic physics community focused on more accurate computational scattering methods going beyond the plane wave Born approximation \cite{Segui2002,boteCalculationsInnershellIonization2008}. For instance, as the name suggests, the plane wave Born approximation assumes the incoming and outgoing electron to be in the form of a plane wave, which is reliable for the fast electron with kinetic energy well above the ionization threshold, but becomes a problem when the electron is slow and noticeably distorted by the presence of the target atom. In this case, the distorted wave Born approximation method is developed to take the finite influence of atomic potential into account. Meanwhile, much of the attention in the electron microscopy community was devoted to dynamical scattering to explain the complex phenomenon in energy-filtered imaging\cite{Kohl1985,Stallknecht1996}, diffraction \cite{Wang1990}, EDS \cite{Rossouw1984} and EELS \cite{Muller1995,Cosgriff2005,Allen2008,Oxley2008,DAlfonso2008,Rusz2017}. In particular, Yoshioka \cite{Yoshioka1957} demonstrated that the inelastic scattering and the associated absorption can be treated as an additional transition potential with an imaginary part, forming the basis for the implementations of dynamic inelastic scattering in various algorithms, including Bloch wave \cite{Maslen1983,Maslen1984,Allen2003}, multislice \cite{Kohl1985,Dwyer2005,Dwyer2008,lobatoProgressNewAdvances2016,majertSimulationAtomicallyResolved2019} and more recently the PRISM algorithm \cite{Brown2019}. The fast electrons used in mid/high voltage electron microscopes experience significant relativistic effects, which anisotropically enhance the cross-sections across energy-momentum space. The influence of relativistic electrodynamics on the differential scattering cross-sections of electron beams was first investigated by Fano in 1956 \cite{fanoDifferentialInelasticScattering1956} and has since been incorporated into various microscopy studies \cite{majertSimulationAtomicallyResolved2019,kurataImportanceRelativisticEffect1997,knippelmeyerRelativisticIonisationCross1997,Schattschneider2005a,Dwyer2006}, though typically using orbital wavefunctions derived from the Schrödinger equation. In contrast, Salvat \cite{Segui2002,boteCalculationsInnershellIonization2008, salvatInelasticCollisionsFast2022} have developed fully relativistic cross-section calculations based on the Dirac equation for atomic and nuclear physics applications. In addition, significant interest has emerged in the DFT community to calculate the fine structure with the density of states explicitly included for EELS \cite{jorissen2010multiple} and X-ray absorption spectroscopy (XAS) \cite{rehr2009ab}, all based on the Schrödinger picture. Given our focus is on the absolute cross-sections for EELS microanalysis, the dynamic scattering effects, and fine structures are beyond the scope of this paper. We noticed a clear pattern in the literature that electron microscopy research typically adheres to the Schrödinger framework, whereas the atomic physics community has long utilized the Dirac framework. Therefore, this paper also aims to bridge the conventions, recognizing that both fields are engaged in the same endeavour.

For a clear comparison with existing GOS databases commonly used within the EELS community, we note that this work was initiated in parallel with the recent Schrödinger GOS database \cite{Segger2019,Segger2023}, which calculates radial wavefunctions using a DFT-based generalized norm-conserving pseudopotential \cite{hamann1989generalized} (hereafter referred to as the DFT-based GOS). Upon learning of this concurrent effort, we adapted our approach to align with the GOSH file standard \cite{GOSH2023}, enhancing accessibility for both users and future developers. Unlike the Schrödinger-based GOS databases (whether HF or DFT-based), we chose to use the Dirac equation due to its intrinsic incorporation of relativistic effects. This approach is particularly important as electron speeds approach the speed of light, especially in heavy elements. It is worth noting that our work is not the first to explore Dirac-based cross-sections in EELS: the widely used GOS database embedded in the commercial Gatan Digital Micrograph software (referred to hereafter as the Gatan GOS) employed the Liberman program \cite{liberman1971} to solve the Dirac equation \cite{Rez2002,Rez2003}. However, the Gatan GOS database did not incorporate the full suite of relativistic effects in cross-section calculations, such as contributions from the small component, variations in spin states, and corrections for relativistic electrodynamics. Additionally, Salvat has developed a Dirac-based GOS (GOSAT) \cite{salvatInelasticCollisionsFast2022} with radial wavefunction solutions calculated from \cite{Salvat2019} for Monte Carlo electron scattering simulations \cite{Salvet2023penelope}. This paper builds upon these foundational efforts by incorporating the current best understanding of relativistic inelastic scattering, with the aim of advancing quantitative EELS research.

To go beyond the limitations of the current GOS databases for accurate EELS quantification, we report the development of an open-source GOS database using the Dirac equation. In this paper, we will introduce the theoretical background of atomic orbital calculations and inelastic electron scattering. First, we will take the reader through the inelastic scattering from the Schrödinger equation in Sec.~\ref{sec:Schrödinger_theory}, followed by a discussion of the relativistic Dirac equation in Sec.~\ref{sec:Dirac_theory} to highlight the connections and differences. In the relativistic inelastic scattering theory, we will consider the relativistic effects on the cross-sections due to fast-moving incident electrons. In Sec.~\ref{sec:geometry}, we will explain the inclusion of experimental parameters of EELS geometry (i.e. convergence angle and collection angle) in the cross-section calculations. Then, we will present the computational details for GOS database generation in Sec.~\ref{sec:computation} followed by the results section in Sec.~\ref{sec:results}. Specifically, Sec.~\ref{sec:overview} will give an overview of the Dirac-based GOS in terms of the energy-momentum sampling compared to existing databases. Then we will explicitly explore the effects of SOC and the contribution of small components which is unique in the Dirac solution. In Sec.~\ref{sec:relativistic_electrodynamics}, we will demonstrate the relativistic electrodynamics effects of the incoming fast electron on the cross-sections for typical acceleration voltages and EELS collection angles. In Sec.~\ref{sec:compare_gos}, we will compare the Dirac-based GOS with the Gatan GOS and the DFT-based GOS for different elemental edges. Sec.~\ref{sec:how_to_use} will briefly demonstrate how to use the Dirac-based GOS database for EELS quantification. Finally, we will conclude the paper in Sec.~\ref{sec:conclusions} with a summary and outlook for future development.

\section{Theory: non-relativistic atomic orbitals and inelastic electron scattering}
\label{sec:Schrödinger_theory}
In this section, we provide an overview of the Schrödinger equation for atomic orbitals and inelastic electron scattering. For a comprehensive treatment of this topic, readers are referred to Cowan's book on atomic spectroscopy \cite{cowan1981theory}, here we only briefly outline the key equations for calculations.

\subsection{Schrödinger equation for atomic orbitals}
\label{sec:Schrödinger_orbital}
In non-relativistic quantum mechanics, the Schrödinger Hamiltonian $\hat{H}_S$ consists of the kinetic energy operator (\(\hat{T}\)) and the potential energy operator (\(\hat{V}\)) for an electron in the presence of a radially symmetric central potential of an atom \(V(r)\) in spherical coordinates:
\begin{equation}
\begin{split}
\hat{H}_S & = \hat{T} + \hat{V} = \frac{\hat{p}^2}{2m_e} -eV(r) \\
& = -\frac{\hbar^2}{2m_e} \frac{1}{r} \frac{\partial}{\partial r^2}r  + \frac{\hat{L}^2}{2m_e r^2} -eV(r)
\label{eq:Schrödinger}
\end{split}
\end{equation}
In this equation, \(\hbar\) is the reduced Planck's constant, \(m_e\) is the mass of the electron, and the linear momentum operator is given by $\hat{p} = -i\hbar\vec{\nabla}$, where $\vec{\nabla}$ is the gradient operator for the wavefunction in space. For a central potential in spherical coordinates, the linear momentum can be further decomposed into a radial part and an angular part. The angular momentum operator $\hat{L}$ is defined as:
    \begin{equation}
        \hat{L^2}=-\hbar^2\left[\frac{1}{\sin \theta} \frac{\partial}{\partial \theta}\left(\sin \theta \frac{\partial}{\partial \theta}\right)+\frac{1}{\sin ^2 \theta} \frac{\partial^2}{\partial \phi^2}\right],
    \end{equation}
with the $\theta$ and $\phi$ being the polar and azimuthal angles, respectively. The projected angular momentum operator $\hat{L}_z$ along the z-axis is given by:
    \begin{equation}
        \hat{L_z} = -i\hbar \frac{\partial}{\partial \phi}.
    \end{equation}
The angular momentum operator $\hat{L} = \vec{r} \times \hat{p}$ commutes with the Hamiltonian, with the eigenvalues of $L^2 =l(l+1)\hbar^2$ and $L_z = m\hbar$ respectively. Here we follow the convention that $n$ is the principal quantum number, $l$ is the orbital angular momentum quantum number, and $m$ is the magnetic quantum number. The eigenfunctions of the Hamiltonian $\psi_{nlm}(r, \theta, \phi)$ can also be separated into a radial part $P_{nl}(r)$ and an angular part $Y_{l}^{m}(\theta, \phi)$:
\begin{equation}
    \psi(r, \theta, \phi) = \frac{1}{r}P_{nl}(r)Y_{l}^{m}(\theta, \phi),
\end{equation}
The angular part $Y_{l}^{m}(\theta, \phi)$ are the well-known spherical harmonics that have analytical expressions. In contrast, the radial wavefunction $P_{nl}(r)$(except for hydrogen) has to be solved numerically  in general with the following ordinary differential equation (ODE):
\begin{equation}
    \frac{d^2 P(r)}{d r^2} = [\frac{l(l+1)}{r^2} - \frac{2m_e}{\hbar^2} (E +e V(r))]P(r).
    \label{eq:Schrödinger_radial}
\end{equation}
The central potential $V(r)$ includes the Coulomb potential from the nucleus and the potential from the other electrons and is spherically averaged:
\begin{equation}
    \begin{split}
        V(r_i) & =  \frac{N e}{r_i} - \sum_{j \ne i}^{N}  \frac{e}{|\vec{r_i} - \vec{r_j}|} \\
        & = \frac{N e}{r_i} - \sum_{j \ne i }^{N}\int \frac{e|\psi_{j}(\vec{r_j})|^2}{|\vec{r_i} - \vec{r_j}|}d \vec{r_j} \\
        & = \frac{N e}{r_i} - \int \frac{e\rho (\vec{r'})}{|\vec{r_i} - \vec{r'}|}d\vec{r'}
    \end{split}
    \label{eq:hartree}
\end{equation}
where $N$ is the positive charge of the nucleus and also the number of total electrons for a neutral atom, $\vec{r_i}$ is the current position of $i$th electron that we want to solve the wavefunction for and $\psi_{j}(\vec{r_j})$ is the wavefunction of the $j$th electron with the position $\vec{r_j}$. To calculate the potential for $i$th electron, we need to sum the Coulomb interaction with all other electrons $j \neq i$. Furthermore, this summation is equivalent to the integral of the charge density $\rho(\vec{r'})$ as a mean field shown in Eq.~\ref{eq:hartree}. One may find the equations are nested in a quite interesting manner: solving the wavefunctions of the current $i$th electron requires the potential in Eq.~\ref{eq:Schrödinger_radial}, which requires solving the wavefunctions and corresponding charge density of all other electrons in Eq.~\ref{eq:hartree}. As a consequence, the solution involves an iterative process known as the self-consistent field (SCF) method which is the key step in the Hartree-Fock-Slater approach or modern density functional theory. For bound states ($E<0$) the radial function is normalized as follows:
\begin{equation}
    \int\psi^{*}(r, \theta, \phi)\psi(r, \theta, \phi) dr=\int_0^\infty P_{nl}^2(r) dr = 1.
\label{eq:Schrödinger_normalization}
\end{equation}
For continuum states ($E>0$), the radial function is normalized by matching the asymptotic form as a plane wave at a large distance away from the nucleus:
\begin{equation}
    P_{El'}(r) \rightarrow \frac{1}{\sqrt{\pi k}} \sin(kr - \frac{\pi}{2}l' -\eta ln(2kr) + \delta_l),
\end{equation}
where $k = \sqrt{2m_e E}/\hbar$ is the wave vector and $\eta$ is the Sommerfeld parameter written as:
\begin{equation}
    \eta = \frac{Z_{\infty}e^2}{\hbar v},
\label{eq:sommerfeld_parameter}
\end{equation}
with $Z_{\infty} = e^{-2}\lim_{r\rightarrow\infty}rV(r)$ which approaches 0 for finite-range potential, and $v$ is the non-realtivistic speed far away from the atom \cite{Salvat2019}. The phase shift $\delta_l$ can be calculated by solving the ODE with the boundary condition $P_{El}(0)=0$ using the Numerov method and matching the asymptotic form at large $r$.

\subsection{Non-relativistic inelastic electron scattering}
Following the derivations given in \cite{schattschneider1986fundamentals}, starting from Fermi's golden rule, the transition rate from initial state $\ket{i}$ to final state $\ket{f}$ can be written as:
\begin{equation}
    \tau_{i\rightarrow f} = \frac{{2\pi}}{{\hbar}} |\bra{f} H'\ket{i}|^2 \rho(E_i,E_f),
\label{eq:fermi_rule}
\end{equation}
where $\tau_{i\rightarrow f}$ represents the transition rate (transition probability per unit time). $\bra{f} H'\ket{i}$ denotes the matrix element (in bra–ket notation) of the interaction Hamiltonian $H'$ between the initial states $\ket{i}$ and final state $\ket{f}$ during the transition. $\rho(E_i, E_f)$ represents the density of states at the initial energy $E_f$ and the final energy. In this study, $\rho(E_i, E_f)$ is taken as a delta function $\rho(E_i, E_f)=\delta(E_i-E_f+\Delta E)$ with $\Delta E = E_f-E_i$ as the energy difference between initial and final states. Since one can distinguish the probe electron and atomic electron (consequently the exchange effects between them are negligible), the initial or final state can be factorized as the wavefunction of the initial and final orbitals $\varphi_i$ and $\varphi_f$ (with N electrons in their atomic reference system $r_1 \ldots r_N$) coupled with the incoming or outgoing plane waves $\Psi_i$ and $\Psi_f$ (in the laboratory reference system $R$), with associated wave vector $\vec{k_i}$ and $\vec{k_f}$:
\begin{equation}
    \begin{split}
        \ket{i} = \ket{\varphi_i} \ket{\Psi_i} = \ket{\varphi_i(\vec{r_1} \ldots \vec{r_N})} \ket{e^{i\vec{k_i}\cdot\vec{R}}},\\
        \ket{f} = \ket{\varphi_f} \ket{\Psi_f} = \ket{\varphi_f(\vec{r_1} \ldots \vec{r_N})} \ket{e^{i\vec{k_f}\cdot\vec{R}}}.         
    \end{split}
\label{eq:initial_final_state}
\end{equation}
To satisfy the anti-symmetry requirement (Pauli principle) upon exchange of two electrons in the many-electron system, one typically uses the Slater determinant to represent the atomic electron wavefunctions  $\varphi_i$ and $\varphi_f$. The Slater determinant is written as:
\begin{equation}
    \ket{\varphi} =\frac{1}{\sqrt{N !}}\left|\begin{array}{cccc}
        \psi_1\left(\vec{r}_1\right) & \psi_2\left(\vec{r}_1\right) & \cdots & \psi_N\left(\vec{r}_1\right) \\
        \psi_1\left(\vec{r}_2\right) & \psi_2\left(\vec{r}_2\right) & \cdots & \psi_N\left(\vec{r}_2\right) \\
        \vdots & \vdots & \ddots & \vdots \\
        \psi_1\left(\vec{r}_N\right) & \psi_2\left(\vec{r}_N\right) & \cdots & \psi_N\left(\vec{r}_N\right)
        \end{array}\right|,
\end{equation}

The interaction Hamiltonian is taken as the Coulombic potential between the probe electron and the target atom similar to Eq.~\ref{eq:hartree}, we can evaluate the transition matrix element by integrating in real space. The real space integral can be simplified using the Fourier transform of the Coulombic potential to the momentum space:
\begin{equation}
    \int d^3 \vec{R} \frac{e^2 }{|\vec{R}-\vec{r_j}|}e^{i\vec{q}\cdot\vec{R}} = \frac{4\pi e^2}{q^2}e^{i\vec{q}\cdot\vec{r_j}},
\label{eq:fourier_coulomb}
\end{equation}
with $\vec{R}-\vec{r_j}$ defined as the vector between the probe electron and the $j$-th atomic electron. After the integration of the laboratory real space $\vec{R}$, the transition matrix element in the momentum space can be evaluated by integration of the atomic reference system for each electron $(\vec{r_1} \ldots \vec{r_N})$ as:
\begin{equation}
\begin{split}
    \bra{f} V \ket{i} & = \frac{e^2}{(2\pi)^2 q^2}(\bra{\varphi_f} \sum_{j}^{N}e^{i\vec{q}\vec{r_j}} \ket{\varphi_i} - N \bra{\varphi_f}\ket{\varphi_i})\\
    & = \frac{e^2}{(2\pi)^2 q^2} \bra{\psi^j_f}e^{i\vec{q}\vec{r_j}}\ket{\psi^j_i}.
\end{split}
\label{eq:transition_matrix}
\end{equation}
Several observations can be made: (1) the plane wave with wave vector of the momentum transfer $\vec{q}=\vec{k_i}-\vec{k_f}$ emerges as a result of the substitution of the explicit form of the initial and final states defined in Eq.~\ref{eq:initial_final_state} into the matrix element in Eq.~\ref{eq:fermi_rule} and Fourier transform of the Coulombic potential in Eq.~\ref{eq:fourier_coulomb}; (2) the contribution to the transition rate from the nuclear electrostatic potential is zero when the initial and final state are orthogonal to each other; (3) only the excited orbital (for instance, the $j$th electron in the equation above) contributes to the transition matrix element due to orthonormality, if we assume all the other orbitals remain unchanged under the frozen core approximation.

The transition rate is dependent on the current density of the quantum system, a more useful definition is the so-called double differential cross-section, which is defined as the probability of the transition per unit solid angle and energy collected. The double differential scattering cross-section (DDSCS) can be written as:
\begin{equation}
    \begin{split}
        \frac{\partial^2 \sigma}{\partial E \partial \Omega} & = (\frac{2\pi}{\hbar})^4 (\gamma m_e)^2 \sum_{i,f} \frac{k_f}{k_i} |\bra{f} V \ket{i}|^2 \delta(E_i-E_f+\Delta E)\\
        & = \frac{4 \gamma^2}{a_0^2 q^4} \frac{k_f}{k_i} |S(q,E)|^2, \\
        & = \frac{4 \gamma^2 }{q^2} \frac{R}{\Delta E} \frac{k_f}{k_i} f(q,E)\\
        \label{eq:ddscs}
    \end{split}
\end{equation}
where $\gamma$ is the Lorentz factor for relativistic kinematics correction, $a_0$ is the Bohr radius, $R$ is the Rydberg energy, $k_i$ and $k_f$ are the wave number of the incoming and outgoing plane waves, respectively. To separate the part that is independent of the incoming electron energy, we define the so-called GOS as $f(q, E)$:
\begin{equation}
    f(q,E) = \frac{\Delta E}{R} \frac{\left|S(q,E)\right|^2}{(qa_0)^2}.
\label{eq:GOS}
\end{equation}
Here $|S(q, E)|^2$ is known as the dynamic form factor, which is the squared magnitude of the transition matrix element summed over the final states as: 
\begin{equation}
    S(q,E) = \sum_{\psi_i, \psi_f}  \bra{\psi_f} e^{i\vec{q}\cdot\vec{r}} \ket{\psi_i}
    \label{eq:dynamic_structure_factor_initial}
\end{equation}
Note that we take the transition matrix only contributed by the excited orbital and the j symbol in Eq.~\ref{eq:transition_matrix} is dropped thereafter for simplicity. The initial and final states of the atomic orbitals are decomposed in terms of radial and angular parts, which has been explained in Sec.~\ref{sec:Schrödinger_orbital}. For the initial states, the summation is applied to the magnetic quantum number $m$ since we have already chosen the orbital. For the final states, the summation extends over both the angular quantum number $l'$ and the magnetic quantum number $m'$ for each specified final state energy $E$.

To derive the dynamic structure factor, one can expand the plane wave using spherical harmonics and integrate the radial and angular parts separately. The plane wave expansion can be written as \cite{olver2010nist}:
\begin{equation}
    e^{iqr} = \sum_{\lambda=0}^{\infty} \sum_{m_\lambda=-\lambda}^{\lambda} 4\pi i^\lambda (2\lambda+1) j_\lambda(qr) Y_{\lambda}^{m_\lambda}(\theta, \phi)
\end{equation}
where $j_\lambda(qr)$ is the spherical Bessel function of the order $\lambda$. First, we can collect all the radial components from the initial states, the final states and the plane wave expansion for the radial integral $R^{\lambda}_{n l, E l'}$:
\begin{equation}
    R^{\lambda}_{nl, El'}(q,E)=\int_0^{\infty}\left[P_{n l}(r) P_{ E l^{\prime}}(r)\right] j_\lambda(q r) d r.
    \label{eq:radial_integral}
\end{equation}
Then we can find the angular integral of the products of three spherical harmonics leads to the Wigner 3j symbol $(^{...}_{...})$, also known as Gaunt coefficients:
\begin{equation}
    \int Y_{\lambda}^{m_{\lambda}}(\theta, \phi) Y_{l^{\prime}}^{m_{l^{\prime}}}(\theta, \phi) Y_{l}^{m_{l}}(\theta, \phi) d \Omega=\sqrt{\frac{(2l+1)(2l'+1)(2\lambda+1)}{4\pi}}\left(\begin{array}{ccc}
        l^{\prime} & \lambda & l \\
        0 & 0 & 0
        \end{array}\right) 
        \left(\begin{array}{ccc}
            l^{\prime} & \lambda & l \\
            m^{\prime} & m_\lambda & m
            \end{array}\right).
\label{eq:gaunt}
\end{equation}
When summing the magnetic quantum numbers $m$ and $m'$ of the initial and final states in Eq.~\ref{eq:gaunt}, the expression can be simplified by applying the orthogonal relation of the Wigner 3j symbols, which results in a delta function and a triangle condition:
\begin{equation}
    \begin{split}
    \sum_{m,m'} \left(\begin{array}{ccc}
        l^{\prime} & \lambda & l \\
        m^{\prime} & m_\lambda & m
        \end{array}\right)\left(\begin{array}{ccc}
        l^{\prime} & \lambda' & l \\
        m^{\prime} & m_\lambda' & m
        \end{array}\right) & = \frac{1}{2\lambda+1}\delta_{\lambda^{\prime}\lambda}\delta_{m_\lambda m_\lambda'}\Delta (l,l',\lambda),\\
        \text{with triangle condition } \Delta \left(l,l',\lambda\right) & = \begin{cases}1 & \text { if }\left|l-l'\right| \leq \lambda \leq l+l' \\ 0 & \text { otherwise }\end{cases}
        \end{split}
\end{equation}
With the radial and angular integral simplified, the dynamic form factor can be written as:
\begin{equation}
    \left|S(q,E)\right|^2= \sum_{l^{\prime}}\sum_\lambda [l,l',\lambda] {R^{\lambda}_{n l, E l'}}^2\left(\begin{array}{ccc}
        l^{\prime} & \lambda & l \\
        0 & 0 & 0
        \end{array}\right)^2,   
\label{eq:dynamic_form_factor} 
\end{equation}
where the abbreviated notion of $[l,l',\lambda] = (2l+1)(2l'+1)(2\lambda+1)$ is used. The dynamic form factor can be physically interpreted as the space-time Fourier transform of the density autocorrelation during the transition, as derived in \cite{schattschneider1986fundamentals}.

The calculations of theoretical spectra, incorporating experimental parameters and Generalized Oscillator Strengths (GOS), will be detailed in Section~\ref{sec:geometry} since the EELS geometric considerations are independent of the choice between Schrödinger or Dirac solutions.

\section{Theory: relativistic atomic orbitals and inelastic electron scattering}
\label{sec:Dirac_theory}
The Dirac equation combines special relativity and quantum mechanics -- the cornerstone of modern physics -- for relativistic electrons \cite{dirac1928quantum}.  In this section, we provide an overview of the Dirac equation for atomic orbitals and inelastic scattering. For a comprehensive treatment of this topic, readers are referred to books on relativistic quantum mechanics \cite{greiner2000relativistic} and recent papers on inelastic electron scattering \cite{majertSimulationAtomicallyResolved2019, Salvat2019} as here we only outline the essential equations.

\subsection{Dirac equations for atomic orbitals}
\label{sec:Dirac_orbital}
The Dirac equation is a first-order differential equation that describes the behaviour of relativistic electrons in the presence of an electromagnetic field (minimum coupling), which is written as:
    \begin{equation}
        (c\gamma^\mu (i\hbar\partial_\mu +eA_\mu) - m_e c^2) \psi = 0,
    \end{equation}
where the summation is implicitly applied over the values of the index $\mu = 0, 1, 2, 3$ in the Einstein summation notation. For instance, the multiplication of two arbitrary 4-vectors $\mathbf{A}$ and $\mathbf{B}$ is written as:
\begin{equation}
    \mathbf{A}\cdot\mathbf{B} = A_\mu B^\mu = A^\nu B_\nu =A_\mu \eta^{\mu \nu} B_\nu =  a^0b^0 -\vec{a}\cdot\vec{b},
\end{equation}
where $A_\mu$ are the covariant 4-components  and  $A^\mu$ are the contravariant 4-components of the vector $\mathbf{A}$ in the tensor index notation, which relates to each other via the Minkowski metric $\eta^{\mu v}$. The multiplication is written in such a manner for the invariance under the Lorentz transformation of the space-time. Back to the Dirac equation, $\partial_\mu = (\frac{1}{c}\frac{\partial}{\partial t}, \vec{\nabla})$ are the 4-gradient for space-time vectors $x^\mu = (ct, \vec{r})$, which gives the 4-momentum $p_\mu = (\mathbb{E}/c,-\hat{p})$ where $\mathbb{E}$ is the relativistic energy of the electron. $A_\mu = (V/c, -\vec{A})$ is the 4-potential consists of electrostatic potential $V$ and magnetic vector field potential $\vec{A}$.  The Dirac gamma matrices $\gamma^\mu$ are defined as:
    \begin{equation}
        \gamma^0 = \begin{pmatrix}
        I_2 & 0 \\
        0 & -I_2
        \end{pmatrix},
        \gamma^i = \begin{pmatrix}
        0 & \sigma_i \\
        -\sigma_i & 0
        \end{pmatrix},
    \end{equation}
where $I_2$ is the 2 $\times$ 2 identity matrix and $\sigma_i$ represents the Pauli matrices:
    \begin{equation}
        \sigma_1 = \begin{pmatrix}
        0 & 1 \\
        1 & 0
        \end{pmatrix},
        \sigma_2 = \begin{pmatrix}
        0 & -i \\
        i & 0
        \end{pmatrix},
        \sigma_3 = \begin{pmatrix}
        1 & 0 \\
        0 & -1
        \end{pmatrix}.
    \end{equation}

We can express the Dirac equation in the 2-component form as follows:
\begin{equation}
    \begin{split} 
        \begin{pmatrix}
            \mathbb{E} + eV(r) - m_ec^2 & -c{\sigma}\cdot(\hat{p}+e\vec{A}) \\
        -c{\sigma}\cdot(\hat{p}+e\vec{A}) & \mathbb{E} + eV(r) + m_ec^2 
        \end{pmatrix}        
        \begin{pmatrix}
        \psi_1 \\
        \psi_2
        \end{pmatrix}
        & =        
        \begin{pmatrix}
            0 \\
            0
        \end{pmatrix},
    \end{split}
    \label{eq:dirac_2_components}
\end{equation}
where the upper component $\psi_1$ and lower component $\psi_2$ are coupled in Eq.~\ref{eq:dirac_2_components}. We may define the energy in the excess of the electron rest energy as $E = \mathbb{E} - m_{e}c^2$ for ease of comparison with the energy in the Schrödinger equation. For an atomic electron in the presence of a radially symmetric electrostatic potential, we may only consider the electrostatic potential $V(r)$ and ignore the magnetic vector field potential $A_\mu$. In the low-energy limit $E + eV(r) << mc^2$, we can see that the lower component $\psi_2$ is much smaller than $\psi_1$ (as $\psi_2 \approx \frac{\sigma \cdot \hat{p}}{2 m_{e}c}\psi_1$), which is at the order of expectation value of the speed relative to the light $\frac{\langle v \rangle}{c}$. For the H atom, this magnitude is at the fine-structure constant ($\sim$ 1/137) but gradually increases with atomic number as the momentum increases. Hence, the upper and lower components are often referred to as the large and small components, respectively. 

The solution can be factorized into radial part and angular parts:
\begin{equation}
    \psi = \frac{1}{r}\begin{pmatrix} 
        P_{n\kappa}(r) \mathscr{Y}_{j l m}(\theta, \phi) \\ 
        iQ_{n\kappa}(r) \mathscr{Y}_{j l^{\prime} m}(\theta, \phi)
        \end{pmatrix},
\label{eq:dirac_solution}
\end{equation}
As $l$ is no longer a good quantum number in the Dirac equation. Here, we introduce the total angular momentum operator defined as $\hat{J}=\hat{L}+\hat{S}$ with its eigenvalue $J^2 =(j+1)j\hbar^2$ and $J_z =m_j\hbar$, and $j$ is bounded by $|l-s| \le j \le |l+s|$ in magnitude. $\hat{S}$ is the spin operator with eigenvalues of $S^2 = s(s+1)\hbar^2$ and $S_z =m_s\hbar$, where $s=1/2$ and $m_s=-1/2$ or $1/2$ for electron. $P_{n\kappa}(r)$ and $Q_{n\kappa}(r)$ are the radial parts of the large and small components, respectively. The  relativistic quantum number $\kappa$ is defined as:
\begin{equation}
    \begin{split}
        \kappa & = (l-j)(2j+1)  =  \begin{cases}  -l-1, & \text { if } j = l + \frac{1}{2} \text{ (spin up)}\\l, & \text { if } j = l -\frac{1}{2} \text{ (spin down)}\end{cases}\\
    \end{split}. 
\end{equation}

For the angular part, $\mathscr{Y}_{j l m}(\theta, \phi)$ represents the spinor spherical harmonics (note in the small component $l'=2j-l$ in Eq.~\ref{eq:dirac_solution}), which can be written as:
\begin{equation}
    \mathscr{Y}_{j l m}(\theta, \phi)=\sum_{m_l, m_{s}}\langle l, 1 / 2, m_l, m_{s} \mid j, m_j\rangle Y_{l}^{m_l}(\theta, \phi) \chi_{\frac{1}{2},m_{s}},
\end{equation}
where $\langle l, 1 / 2, m_l, m_s \mid j, m_j\rangle$ is the Clebsch-Gordan coefficient, $Y_{l}^{m_l}(\theta, \phi)$ is the usual spherical harmonics with $m_l=m_j-m_s$ and $\chi_{\frac{1}{2},m_s}$ is the spin eigenfunction of a spin 1/2 particle:
\begin{equation}
    \chi_{\frac{1}{2},+\frac{1}{2}}=\left(\begin{array}{c}
        1 \\
        0
        \end{array}\right) \quad \text { and } \quad \chi_{\frac{1}{2},-\frac{1}{2}}=\left(\begin{array}{c}
        0 \\
        1 \end{array}\right)
\end{equation}

%

For the radial part, the Dirac equation can be written as coupled ODE equations:
\begin{equation}
    \begin{split}
        \frac{\mathrm{d} P_{n \kappa}}{\mathrm{d} r} & =-\frac{\kappa}{r} P_{n \kappa}+\frac{E+2 \mathrm{~m}_{\mathrm{e}} c^2+eV}{c \hbar} Q_{n \kappa}, \\
        \frac{\mathrm{d} Q_{n \kappa}}{\mathrm{d} r} & =\frac{-E-eV}{c \hbar} P_{n \kappa}+\frac{\kappa}{r} Q_{n \kappa}.
    \end{split}
    \label{eq:Dirac_radial}
\end{equation}
For the bound state ($E<0$), the radial part can be normalized as:
\begin{equation}
    \int_0^{\infty} \left(P_{n \kappa}^2(r)+Q_{n \kappa}^2(r)\right) \mathrm{d} r=1.
    \label{eq:dirac_normalized}
\end{equation}
For the free state ($E>0$), the radial part can be normalized by matching the asymptotic behaviour:
\begin{equation}
    P_{E \kappa'} \rightarrow \frac{1}{\sqrt{\pi k}}\sin \left(k r-\frac{\pi}{2}\kappa' -\eta ln(2kr) +\delta_{\kappa'}\right),
\end{equation}
where $k = \sqrt{E(E+2m_ec^2)}/\hbar c$ is the relativistic wave number, and $\delta_{\kappa'}$ is the phase shift. $\eta$ is the relativistic Sommerfeld parameter with the corresponding relativistic speed in Eq.~\ref{eq:sommerfeld_parameter}. In the low-energy limit, $E+eV<<m_e c^2$, the radial part can be written as \cite{Salvat2019}:
\begin{equation}
        Q_{n\kappa}=\frac{c \hbar}{2 \mathrm{~m}_{\mathrm{e}} c^2}\left(\frac{\kappa}{r} P+\frac{\mathrm{d} P}{\mathrm{~d} r}\right)
\end{equation}
and 
    \begin{equation}
        \frac{d^2 P_{n\kappa}(r)}{d r^2} = [\frac{\kappa(\kappa+1)}{r^2} - \frac{2m_e}{\hbar^2} (E +eV(r))]P_{n\kappa}(r)
    \label{eq:dirac_radial_low_energy}
    \end{equation}
Note that $\kappa(\kappa+1)=l(l+1)$, reducing the upper component $P_{n\kappa}(r)$ radial equation to the Schr\"{o}dinger radial equation as in Eq.~\ref{eq:Schrödinger_radial}.

\subsection{Relativistic inelastic electron scattering}
From the perspective of electrodynamics, the motion of a charged particle induces electric and magnetic fields and hence perturbs the interaction Hamiltonian. Instead of taking the rigorous quantum field theory, here we follow perturbation theory due to its simplicity, as well outlined in \cite{Segui2002, knippelmeyerRelativisticIonisationCross1997,Schattschneider2005a,Dwyer2006}. The DDSCS for inelastic electron scattering is given by:
\begin{equation}
    \frac{\partial^2 \sigma}{\partial E \partial \Omega}=  {\left(\frac{2 \gamma }{a_0}\right)^2 \frac{1}{\left(q^2- (\Delta E/\hbar c)^2\right)^2} \frac{k_f}{k_i} \sum_{\psi_i,\psi_f} \mid\langle \psi_f| e^{i \vec{q} {\vec{r}}}\left(1-\frac{\hat{p} \vec{v}_0}{m_e c^2}\right) }\left.|\psi_i\rangle\right|^ 2 \delta\left(E_i-E_f+\Delta E\right),
    \label{eq:DDSCS_relativistic}
\end{equation}
where $v_0$ is the velocity of the incident probe electron with energy $E_0$. By comparing Eq.\ref{eq:DDSCS_relativistic} with the non-relativistic case in Eq.~\ref{eq:ddscs}, we can note that $q^2$ is replaced by $q^2- (\Delta E/\hbar c)^2$ and there is an extra term $\frac{\hat{p} \vec{v}_0}{m_e c^2}$, which originates from the perturbation of the fast moving charge under the Coulomb gauge leading to the contraction of the scattering vector in the direction of the incident beam (retardation effect) \cite{Schattschneider2005a}. This is equivalent to the z-component of the 4-transition current under the Lorentz gauge \cite{Dwyer2006}. The relativistic effect of incident fast electron plays a critical role in the experimental observation of the magic angle in EELS \cite{Schattschneider2005a} and the detailed intensity distribution in energy-filtered electron diffraction \cite{Dwyer2006}, thus it is important to include it correctly for quantitative work. By projecting the contribution of $\frac{\hat{p} \vec{v}_0}{m_e c^2}$ to directions that are parallel and normal to the incident electron, we can further simplify the DDSCS as proposed in \cite{Segui2002,knippelmeyerRelativisticIonisationCross1997}:
\begin{equation}
    \frac{\partial \sigma}{\partial E \partial \Omega} =\left(\frac{2 \gamma}{a_0}\right)^2 \frac{k_f}{k_i} \left[ \frac{1}{q^4} + \frac{\beta_t ^2 (\Delta E/\hbar c)^2}{(q^2-(\Delta E/\hbar c)^2)^2} \right]  \sum_{\psi_i, \psi_f}|\langle \psi_f|e^{i\vec{q}\vec{r}}| \psi_i\rangle|^2,
    \label{eq:DDSCS_relativistic_simplified}
\end{equation}
where $\beta_t$ is the transverse component of the incident electron velocity (in units of the speed of light) $\beta = v_0/c$ with respect to the scattering vector $\vec{q}$:
\begin{equation}
    \beta^2_t = \beta^2 - \frac{\Delta E^2}{(cq)^2}(1+\frac{(cq)^2 - \Delta E^2}{2\Delta E(E_0+m_e c^2)})
\end{equation}

For the orbital relativistic effects, the transition matrix element must account for the Dirac solution, which includes both the large and small components \cite{Segui2002}. As a consequence, the radial overlap integral is changed from Eq.~\ref{eq:radial_integral} to:
    \begin{equation}
        R^{\lambda}_{n \kappa, E \kappa'}(q,E)=\int_0^{\infty}[P_{n \kappa}(r) P_{ E \kappa^{\prime}}(r) + Q_{n \kappa}(r) Q_{ E \kappa^{\prime}}(r)] j_\lambda(q r) d r.
        \label{eq:radial_integral_relativistic}
    \end{equation}
    For the angular part, the spherical harmonics in the Schrödinger solution are now replaced by the spinor spherical harmonics in the Dirac solution. This addition introduces a Wigner 6j $\{_{...}^{...}\}$ term when integrating over the angle and summing over the states. Overall, the dynamic form factor in the Dirac solution is given by:
    \begin{equation}
        \left|S(q,E)\right|^2= \sum_{l^{\prime}}\sum_\lambda \sum_{\kappa'}[l,l',j',\lambda] {R^{\lambda}_{n \kappa, E \kappa'}}^2\left(\begin{array}{ccc}
            l^{\prime} & \lambda & l \\
            0 & 0 & 0
            \end{array}\right)^2 \left\{\begin{array}{ccc}
                j & \lambda & j' \\
                l' & \frac{1}{2} & l
                \end{array}\right\}^2,   
    \label{eq:dynamic_form_factor_relativistic} 
    \end{equation}
    The GOS still follows the same definition as in the non-relativistic case in Eq.~\ref{eq:GOS}.

To check we have the correct implementation of the relativistic correction defined in Eq.~\ref{eq:DDSCS_relativistic_simplified}, we can use the analytical expression for the relativistic correction ratio under dipole approximation, which is defined as the ratio of the scattering cross-sections with and without the relativistic electrodynamics effects. This ratio is given by \cite{majertSimulationAtomicallyResolved2019,Schattschneider2005a,Dwyer2006}:
\begin{equation}
    \sigma_{\text {ratio }}(\Delta E)=\frac{\sigma_{\mathrm{rel}}(\Delta E)}{\sigma_{\text {conv }}(\Delta E)}=\left[\ln \left(1+\frac{x_\theta}{1-\beta^2}\right)-\frac{\beta^2 x_\theta}{1-\beta^2+x_\theta}\right] \frac{1}{\ln \left(1+x_\theta\right)},
    \label{eq:relativistic_dipole_correction}
\end{equation}
where $x_\theta = \theta_0^2/\theta^2_E$ is the squared ratio of the collection angle $\theta_0$ to the characteristic angle $\theta_E$, which is written as:
\begin{equation}
   \theta_E = \frac{\Delta E}{\gamma mv_0^2}
\end{equation}
In Section~\ref{sec:relativistic_electrodynamics}, we will examine the relativistic correction ratio for cross-sections under different experimental conditions.

\section{EELS geometry parameters in momentum space integration}
\label{sec:geometry}

In this section, we will explain how to consider the experimental EELS geometry (i.e. EELS collection angle and STEM convergence angle) for cross-section calculations. From the dynamic form factor in Eq.~\ref{eq:dynamic_form_factor} and Eq.~\ref{eq:dynamic_form_factor_relativistic} (or the associated GOS in Eq.~\ref{eq:GOS}), one can integrate over the EELS collection angle to obtain the differential cross section as a function of energy loss \cite{schattschneider1986fundamentals}. The differential cross section is given by:
    \begin{equation}
        \frac{d\sigma}{dE} = \frac{4\pi\gamma^2}{k_i^2} \int^{Q_{\text{max}}}_{Q_{\text{min}}}\frac{|S(q,E)|^2}{Q} d(ln(Q)),
    \label{eq:diff_cross_section_integral_q}
    \end{equation}
where we take $Q=(a_0q)^2$ for ease of integration. From the scattering geometry, we can set the upper and lower limits of the integration as:
    \begin{equation}
        \begin{split}
            q_{\text{min}} = k_i - k_f, \\
            q_{\text{max}}
            = \sqrt{{k_i}^2 + {k_f}^2 - 2{k_i}{k_f}\cos(\beta_0)},\\
        \end{split}
    \end{equation}
as the minimum scattering angle is zero and the maximum scattering angle is bound by the EELS collection aperture $\beta_0$. The differential cross section as a function of energy loss is used to fit and quantify the experimental EELS spectra.

For plane wave illumination in TEM, The above Eq.~\ref{eq:diff_cross_section_integral_q} is sufficient. For STEM convergent beam illumination, we need to take into account the finite angle of both the convergence angle and collection angle, which can be performed by the geometric correction procedure proposed by Kohl \cite{kohl1985simple}. In this approach, we consider the STEM illumination as a collection of plane waves with their corresponding wave vectors bounded by the convergence angle $\alpha_0$. The EELS collection angle $\beta_0$ defines the maximum angle relative to the incident direction to be recorded by the detector. For a given scattering angle $\theta$, there could be different combinations of incident angle and exit angles allowed by the geometry. As a consequence, the DDSCS should be multiplied by a correction factor corresponding to those combinations. Based on the geometric interpretation, this problem is equivalent to the intersection area of two circles -- one with a radius of $\alpha_0$ and another with a radius of $\beta_0$ and they are separated by a distance of $\theta$. The resulting correction factor is a cross-correlation function, which is written as \cite{kohl1985simple}:
\begin{equation}
    \begin{aligned}
        & F_{\mathrm{BF}}\left(\alpha_0,\beta_0,\theta\right) \\
        & \quad=\left\{\begin{array}{l}
        \theta_{<}^2 / \alpha_0^2, \quad \text { for } 0 \leqslant \theta \leqslant\left|\alpha_0-\beta_0\right|, \\
        \pi^{-1}\left[\arccos (x)+\left(\beta_0^2 / \alpha_0^2\right) \arccos (y) \right. \\
        \left. -\left(1 / 2 \alpha_0^2\right) \sqrt{4 \alpha_0^2 \beta_0^2-\left(\alpha_0^2+\beta_0^2-\theta^2\right)^2}\right], \text { for }\left|\alpha_0-\beta_0\right|<\theta<\alpha_0+\beta_0, \\
        0, \text { otherwise. }
        \end{array}\right.
    \end{aligned}
\end{equation}

where 
\begin{equation}
    \begin{split}
    \theta_{<} & =\min \left(\alpha_0,\beta_0\right),\\
    x & =\frac{\alpha_0^2+\theta^2-\beta_0^2}{2 \alpha_0 \theta},\\
    y & =\frac{\beta_0^2+\theta^2-\alpha_0^2}{2 \beta_0 \theta}.
    \end{split}
\end{equation}
The EELS correction factor $F_{BF}$ (closely related to the contrast transfer function of bright field incoherent imaging) is used to correct the DDSCS for a given set of scattering angles in the STEM-EELS geometry. The effective partial scattering cross-section after integration in the momentum space is written as \cite{kohl1985simple}:
\begin{equation}
    \frac{d\sigma_{\text{eff}}}{dE} = \int_0^{\alpha_0+\beta_0}  F_{BF}(\alpha_0,\beta_0,\theta) \frac{d^2\sigma}{dEd\Omega} 2\pi\theta d\theta,
\end{equation}
or in terms of the Q space integral as in Eq.~\ref{eq:diff_cross_section_integral_q}
\begin{equation}
    \begin{split}
        \frac{d\sigma_{\text{eff}}}{dE} & = \frac{4\pi\gamma^2}{k_i^2} \int^{Q_{\text{max}}}_{Q_{\text{min}}}\frac{|S(q,E)|^2}{Q} F_{BF}(\alpha_0,\beta_0,\theta) d(ln(Q)),\\
        \text{ with } \theta & = \arccos\left(\frac{{k_i}^2+{k_f}^2-{q}^2}{2 {k_i} {k_f}}\right),\\
        q_{\text{min}} & = {k_i} - {k_f},\\
        q_{\text{max}} & = \sqrt{{k_i}^2 + {k_f}^2 - 2{k_i}{k_f}\cos(\alpha_0+\beta_0)},
    \end{split}
\label{eq:diff_cross_section_integral_q_BF}
\end{equation}

\section{Computational details}
\label{sec:computation}
In this study, we perform GOS calculations for single atoms only, thereby disregarding the solid state effects in EELS. This is a reasonable approximation as previous studies have shown that the error introduced by solid-state effects is less than 5\% within a reasonable energy window \cite{wengSolid1988}. The transition to discrete unoccupied states is not included in this study. Previous studies have shown that the associated white lines could contribute up to 10\% to the cross sections for 3d transition elements \cite{rezContribution1992}. This contribution will be incorporated in future updates to this database. The atomic orbital calculations are performed using the Flexible Atomic Code (FAC) \cite{Gu2008} package. To solve the Dirac radial wavefunction as described in Eq.~\ref{eq:Dirac_radial}, FAC considers the local atomic potential as a combination of the electron-nucleus and electron-electron Coulomb interactions. The solution is computed through a self-consistent Dirac–Fock–Slater method within a modified local density approximation for the correct asymptotic behaviour of the exchange energy \cite{Gu2008}. We take the bound state of the target orbital as the initial state and the continuum state as the excited state. Continuum states are treated in the distorted-wave approximation \cite{Gu2008}. Fermi's golden rule employed a delta function to represent the density of states, thus ignoring the fine structures. In addition, the orthogonality of the initial and final states is necessary to exclude the contribution of the nucleus in Eq.~\ref{eq:transition_matrix}, which otherwise causes a divergence of the cross-sections when momentum transfer approaches zero. To ensure orthogonality, we take the frozen core approximation that assumes the potential remains unchanged for the initial and final states, which means only the exited orbital contributes to the transition matrix in Eq.~\ref{eq:transition_matrix}. The calculated large and small components are used in Eq.~\ref{eq:radial_integral_relativistic} to compute the transition matrix element. The convergence is checked by including final states (summation of $\kappa'$ in Eq.~\ref{eq:dynamic_form_factor_relativistic}) until the contribution of the last final state falls below 0.1\%. 

The GOS calculation is computationally intensive since we need to sum over different final states for each energy loss. To accelerate the process, we developed a highly efficient parallelization scheme to take advantage of modern computers with multiple CPUs.  Specifically, the calculation of the transition matrix is evaluated using the multi-processing of different energy losses. For a given energy loss, multi-threading is used for all the final states $\kappa'$ until convergence. We implemented a dynamic predicting, caching, and dumping strategy for the number of final state wavefunctions needed to manage memory usage during parallelization. After optimization, the computation of the entire database takes about 2 days on a desktop (Intel i9-7900X CPU 10 cores @4.5GHz), and around 4 hours on a single node of a high-performance computing cluster (AMD Epyc 7H12 CPU 2x64-core @2.6 GHz).

We constructed the GOS database for all elements with atomic number Z ranging from 1 to 118 and all available shells. This resulted in a total of 2143 entries tabulated as a function of energy loss $\Delta E$ and momentum transfer $q$ with fine sampling. More specifically, we computed the GOS in an energy range of 0.01 - 4000~eV above the ionization energy with 128 sampling points and an adaptive momentum range of ($q_{\text{min}}$, $q_{\text{max}}$) with 256 sampling points. Both energy and momentum sampling follow a log sampling scheme, as also done in \cite{Manson1972,Leapman1980a,Segger2023}. The minimum momentum transfer $q_{\text{min}}$ is in the direction of the incident beam, defined as:
\begin{equation}
    q_{\text{min}} = \frac{\Delta E_{\text{min}}}{\hbar c},
    \label{eq:q_min}
\end{equation}
where $\Delta E_{min}$ corresponds to the minimum energy loss which is the ionization energy. The maximum momentum sampling $q_{\text{max}}=2q_{r}$, where $q_{\text{r}}$ is the momentum transfer at the Bethe ridge for a given energy loss \cite{egerton2011electron}:
\begin{equation}
    (a_{0} q_{\text{r}})^2  = \frac{\Delta E}{R} + \frac{\Delta E^2}{2 m_{e} c^2 R}.
    \label{eq:bethe_ridge}
\end{equation}
Adaptive momentum sampling is developed in such a manner to maximize the physical information for a given finite number of sampling points. For instance, an ionization event is always associated with a scattering vector larger than $q_{\text{min}}$, thus sampling below the lower limit would be pointless. Also, the transition probability beyond $q_{\text{max}}$ is well decayed to negligible. Note that the ($q_{\text{min}}$, $q_{\text{max}}$) is elemental edge dependent, so the sampling has to be adaptive instead of fixed for all the edges. The database was saved in an HDF5 file following the GOSH file format \cite{GOSH2023} so that users can switch easily between different variations of GOS databases.

We also used the GPAW software to calculate the all-electron Schrödinger DFT using the projector augmented wave method \cite{blochl2003projector} with generalized gradient approximation of the potential \cite{Perdew1996}. The DFT calculations are performed to compare the orbital energies with Dirac results in this work but not for GOS computation. 

\section{Results and discussions}
\label{sec:results}
In this section, we will first give an overview of the Dirac-based GOS database compared to other Schrödinger solutions in Sec.~\ref{sec:overview}, highlighting the effects of SOC on the ionization energy, the wavefunctions, and the GOS. In Sec.~\ref{sec:relativistic_electrodynamics}, we will demonstrate the necessity of considering the relativistic effects of fast electrons in quantitative EELS. In Sec.~\ref{sec:compare_gos}, we will compare the computed scattering cross-sections from the Dirac-based GOS with other commonly used GOS databases, showing the general agreement and fine differences. In Sec.~\ref{sec:how_to_use}, we will provide a step-by-step guide on how to use the Dirac-based GOS database for EELS quantification.
\subsection{Overview of the Dirac-based GOS database}
\label{sec:overview}
\begin{figure}[htbp]
    \centering
    \includegraphics[width= 1.0\textwidth]{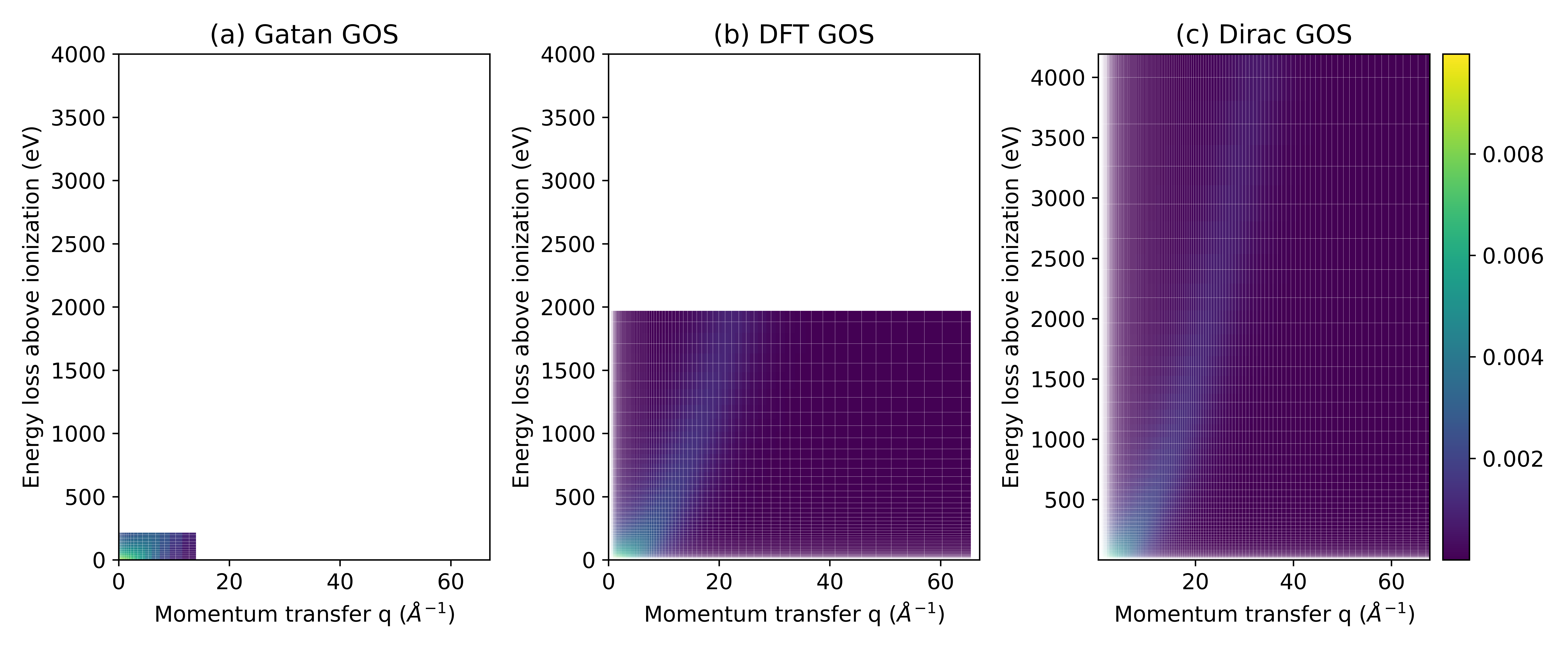}
    \caption{Comparison of the GOS for the C K-edge calculated from (a) the Gatan solution, (b) the DFT solution, and (c) our Dirac solution as a function of momentum transfer and energy loss above the ionization. The sampling grid is indicated by the white lines. The log sampling is dense at low energy loss and low momentum transfer region, where most experiments focus on.}
    \label{fig:gos_sampling}
\end{figure}

A distinct improvement of our GOS database is the large energy-momentum range with a fine sampling. Such sampling requires significant computational efforts, which once limited the Gatan GOS, are now more manageable owing to the rapid development of computation hardware and highly optimized parallelization algorithms. For comparison, Fig.~\ref{fig:gos_sampling} shows the GOS database for the C-K edge constructed previously based on (a) the Gatan Dirac solution, (b) the DFT solution, and (c) our Dirac solution. For this particular case, Gatan GOS reaches a maximum energy of 200~eV above the onset with 12 sampling points and a maximum momentum of 12.7 $\angstrom^{-1}$ with 20 sampling points. For conventional cross-section integration, one may argue that 200~eV should be sufficient for an integration energy window but not enough for model-based fitting. In contrast, the Dirac GOS extends the sampling to 4000~eV above the ionization and 67 $\angstrom^{-1}$ with 128 and 256 sampling points, respectively. The sampling scheme enables the clear visualization of the Bethe ridge with locally high transition probabilities in a parabolic region ($\Delta E \sim q^2$) in the energy-momentum space. In practice, the extended energy loss range of GOS can be useful for the simultaneous model-based fitting of multiple elements. For example, the presence of carbon support is common in TEM sample preparation. In the case of a limited GOS energy range, the background fitting may only use the pre-edge region to extrapolate a power law function instead of taking the exact decaying tail of the C K-edge, which introduces additional bias for model-based fitting of multiple elements. Fine sampling is also helpful for the accurate generation of the EELS cross-sections and consequently spectrum quantification. Note that the large sampling range and fine sampling step is a common practice in modern computing, which is also available at recent DFT-based GOS \cite{Segger2023} shown in Fig.~\ref{fig:gos_sampling} (b) but with a momentum sampling fixed around 63 $\angstrom^{-1}$ and a smaller energy range (up to $\sim$ 1900 eV) for all elements. 

\begin{table}[htbp]
    \centering
    \caption{The onset energies of different edges, with the experimental measurement from EELS Atlas, XPS, and the theoretical predictions from Dirac and all-electron Schrödinger solutions.}
    \label{tab:edge_onset}
    \begin{tabular}{lllcccc}
    \toprule
    \textbf{Element} & \textbf{Orbital} & \textbf{Edge} & \multicolumn{2}{c}{\textbf{Experimental Measurement (eV)}} & \multicolumn{2}{c}{\textbf{Theoretical Calculation (eV)}} \\ 
    \cmidrule(lr){4-5} \cmidrule(lr){6-7}
    & & & \textbf{EELS Atlas}\cite{ahn2006transmission} & \textbf{XPS}\cite{lee2002development} & \textbf{Dirac} & \textbf{Schrödinger} \\ 
    \midrule
    \multirow{2}{*}{Ti} & \ce{2p_{1/2}} & \ce{L2}  & 462 & 460 & 454 & \multirow{2}{*}{444}  \\
                        & \ce{2p_{3/2}} & \ce{L3}  & 456 & 454 & 448 &               \\ 
    \addlinespace
    \multirow{2}{*}{Ag}& \ce{3d_{3/2}} & \ce{M4} & 373 & 374 & 370 & \multirow{2}{*}{358}\\
                        & \ce{3d_{5/2}} & \ce{M5} & 367 & 368 & 364 & \\
    \addlinespace
    \multirow{2}{*}{Au}& \ce{4f_{5/2}} & \ce{N6} & 86 & 88 & 84 & \multirow{2}{*}{81}\\
                        & \ce{4f_{7/2}} & \ce{N7} & 83 & 84 & 80 & \\ 
    \bottomrule
    \end{tabular}
\end{table}

\begin{figure}[htbp]
    \centering
    \includegraphics[width= 1.0\textwidth]{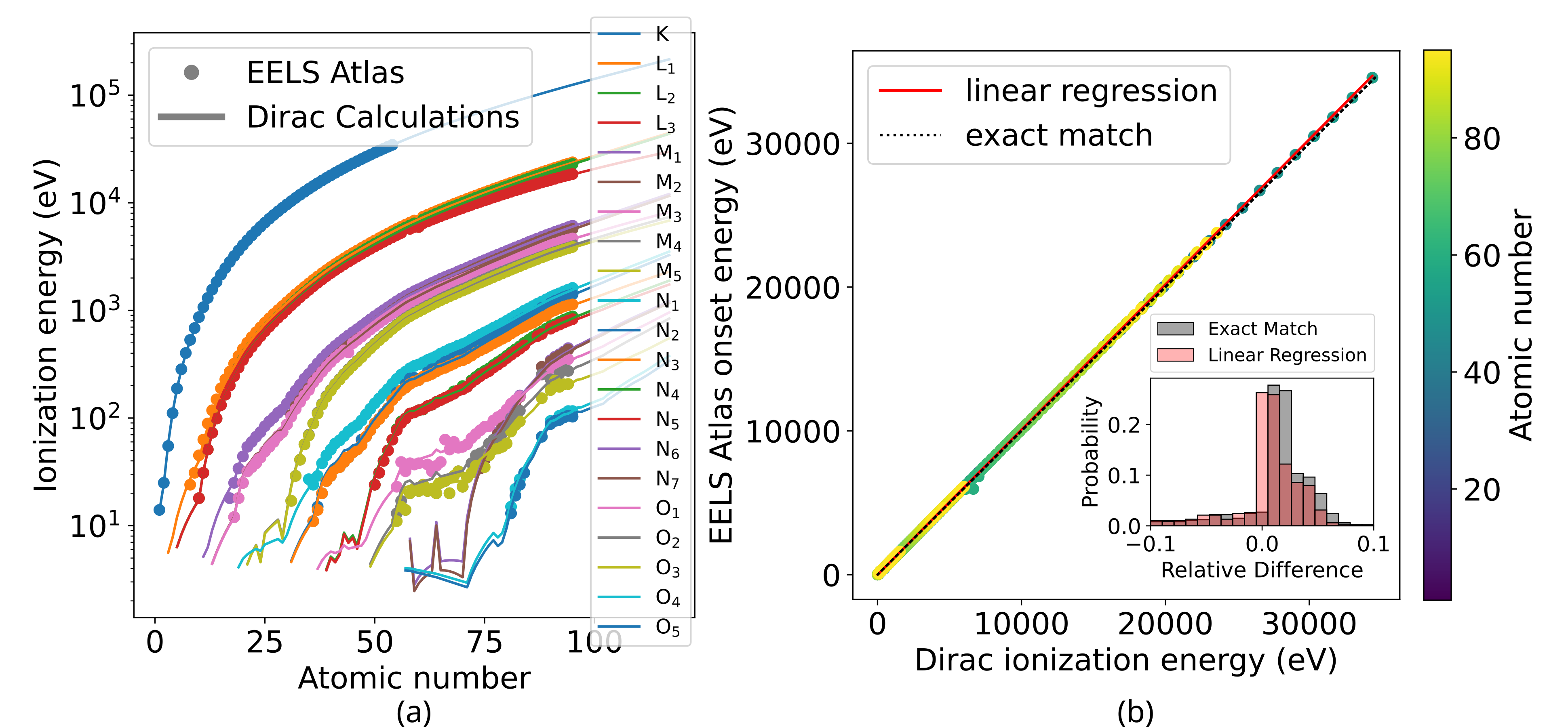}
    \caption{ (a) Comparison of the Dirac predicted ionization energy with the EELS Atlas for all available elements (Z=1-96). (b) A plot of the predicted ionization energies against the edge onset energies from EELS Atlas to demonstrate the linear correlation. The variation of atomic number can be visualized by the colour bar. The black dashed line indicates the exact match between theory and experiment. The red solid line represents the linear regression of the data. The probability histogram of the relative residual between the Dirac calculation and EELS Atlas is plotted inside for the exact match and after linear regression.}
    \label{fig:ionisation_energy}
\end{figure}

The key benefit of employing the Dirac equation lies in its intrinsic inclusion of relativistic effects, rather than applying corrections to the Schrödinger equation. In the context of EELS, a notable advantage is the precise prediction of spin-orbit splittings of different elements. Table~\ref{tab:edge_onset} presents Ti \ce{L2}/\ce{L3}, Ag \ce{M4}/\ce{M5}, and Au \ce{N6}/\ce{N7} edges with experimental measurements from the EELS Atlas \cite{ahn2006transmission}, X-ray Photoelectron Spectroscopy (XPS) \cite{lee2002development} and theoretical ionization energies calculated using the Dirac and Schrödinger equations for each orbital. For instance, the  Ti \ce{L2} edge is observed to be 6~eV higher than the \ce{L3} edge in both EELS and XPS, with the splitting known due to spin-orbit coupling (SOC) \cite{cowan1981theory}. From the perspective of applying relativistic corrections to the Schrödinger equation, a spin-down electron has lower energy than that of a spin-up electron with a SOC contribution that scales with $\sim \langle {\hat{L}\cdot\hat{S}} \rangle$. Such measurements align well with predictions from the Dirac equation (454 / 448~eV). In contrast, the Schrödinger equation predicts a single onset energy for the 2p orbital regardless of spin variations. However, it is important to note that the Dirac values do not exactly match experimental measurements. This discrepancy arises because we only performed the calculation for a single neutral atom, thereby ignoring chemical shifts (as observed in ions with different oxidation states, for example, the Ti edges were measured from \ce{TiO2} in the EELS Atlas) and solid state effects (resulting from many-body perturbations by surrounding atomic potentials). Nevertheless, the Dirac equation accurately predicts the energy difference from splitting as observed experimentally. This trend is similarly observed in the \ce{M4}/\ce{M5} edge of Ag and \ce{N6}/\ce{N7} edge of Au. Note that we did not yet consider the effect of Zeeman splitting of the strong magnetic field in the objective lens of a modern microscope, which could be interesting for further study. To further verify the match between theory and experimental measurements systematically, Fig.~\ref{fig:ionisation_energy}(a) plots the Dirac predicted ionization energy against the edge onset energy all available in EELS Atlas, containing 1105 pairs of data in total and covering a wide range of atomic numbers (1-96) and energies (11-34561~eV), highlighting the overall agreement for different orbitals. To separate the multiple lines, Fig.S1-2 in the supplementary materials provide additional plots for each orbital for a clearer view. As discrepancies between atomic calculations and experimental measurement are expected due to the aforementioned chemical shifts and solid-state effects, small deviations from the exact match (as indicated by the dashed line) are observed in Fig.~\ref{fig:ionisation_energy}(b). A histogram of the relative residual between the Dirac predictions and EELS Atlas values is plotted inside of Fig.~\ref{fig:ionisation_energy}(b), showing the deviations (mean: 0.2\%, standard deviation: 7\%) are well bounded. Further investigation shows that those cases with large deviations are mainly from the O edges of heavy elements (i.e. transition elements in the La and Ac families), which are also visible in Fig.~\ref{fig:ionisation_energy}(a) and Fig.S2 in the supplementary materials. The general trend of residual can be captured fairly well with linear regression as indicated by the red solid line. We can also make a reasonable estimation for other edges not recorded in the EELS Atlas by applying the same slope ($\sim$ 1.01) from the linear regression to the Dirac ionization energy. 

\begin{figure}[htbp]
    \centering
    \includegraphics[width= 0.8\textwidth]{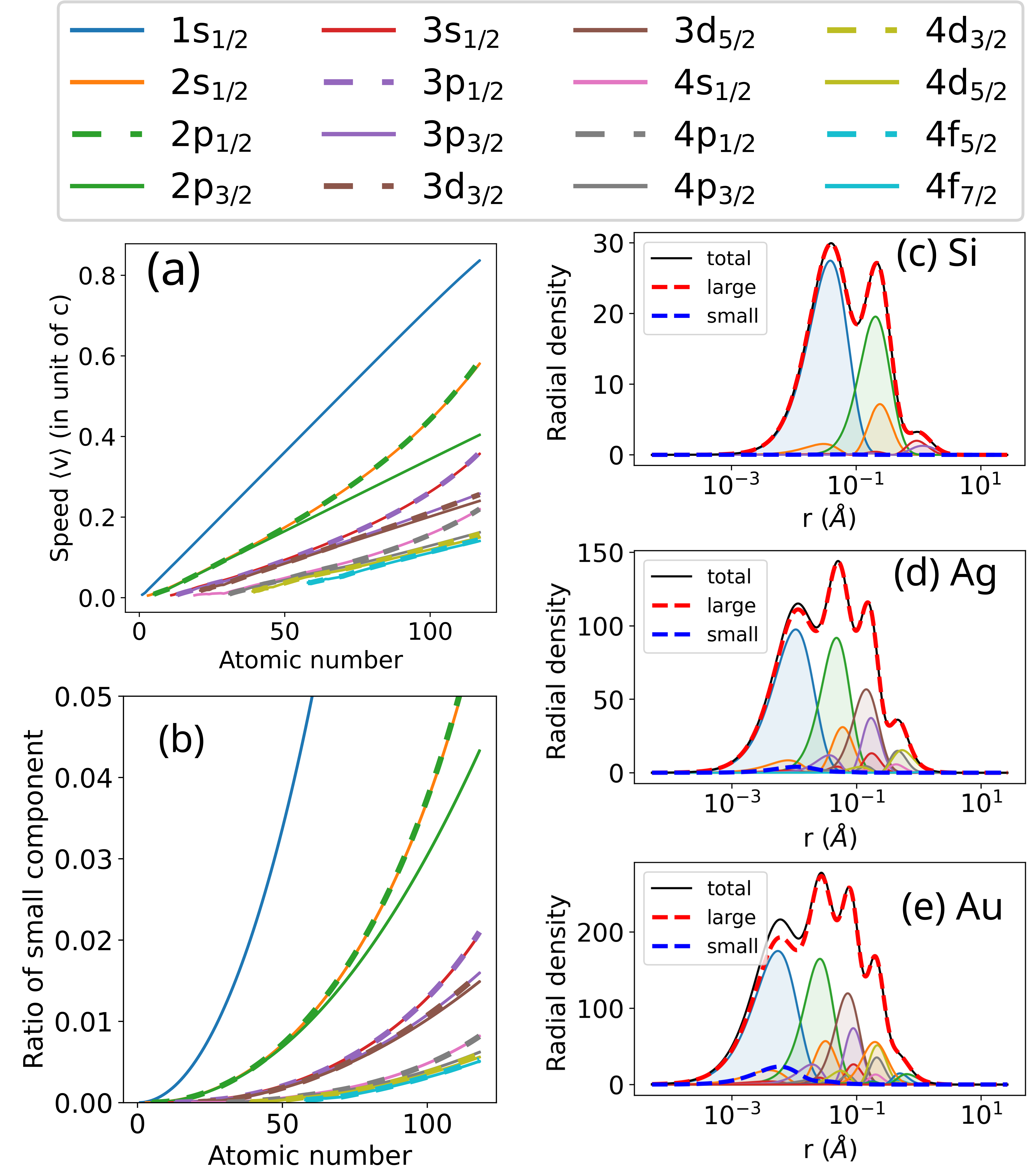}
    \caption{(a) The expectation value of relativistic speed $\langle v \rangle$ of the orbital electrons (up to 4g) for elements (Z=1-118) in the periodic table. The spin-down electron is indicated by the dotted line. Fig.S3 in the supplementary materials provides additional plots for each orbital. (b) Contribution of the small component to the charge density for each orbital of all elements. Fig.S4 in the supplementary materials also presents the results for each orbital. (c-e) The radial charge density for three typical elements from light to heavy (Si, Ag, and Au). The large component is shown in blue and the small component is shown in red. The orbital decomposition is indicated by the shaded area in different colours.}
    \label{fig:orbital_speed}
\end{figure}

In addition to the precise onset energy, the Dirac solution is expected to capture the accurate charge distribution due to relativity. The Dirac solution has large and small components, where the large component is much more significant and can be approximated to the Schrödinger solution at the low-speed limit as shown in Eq.~\ref{eq:Schrödinger_radial} and Eq.~\ref{eq:dirac_radial_low_energy}. However, this does not mean the small component is not important, especially when the core-shell electron of heavy elements approaches the speed of light. The expectation value of the speed of an orbital electron $\langle v \rangle$ is defined as:
\begin{equation}
    \langle v \rangle = \frac{ \langle p \rangle c^2}{\sqrt{m_e^2c^4+\langle p^2 \rangle c^2}}.
\end{equation}
Fig.~\ref{fig:orbital_speed}(a) shows that orbital electrons (up to 4g orbital) $\langle v \rangle$ increase with the atomic number, reaching a considerable fraction of the speed of light for heavy elements. For instance, the 1s electron of Au can reach $\sim$~57\% of the speed of light. Ranking the speed among different orbitals, one may observe that the \ce{1s_{1/2}} electron is always associated with the highest speed as it directly faces the nucleus (with the lowest expectation value of the distance from nucleus $\langle r \rangle$ and therefore least screening effects caused by other electrons). This is then followed by the \ce{2s_{1/2}} and \ce{2p_{1/2}} spin-down electrons with similar velocities. The \ce{2p_{1/2}} electron has a noticeably lower speed compared to that of \ce{2p_{3/2}}. There are additional orbital pairs with similar speeds for \ce{3s_{1/2}}-\ce{3p_{1/2}}, \ce{3p_{3/2}}-\ce{3d_{3/2}}, etc. This is not surprising as such pairs have the same total angular momentum $j$ and relativistic quantum number term $(\kappa+1)\kappa$. Consequently, they have the same effective potential and kinetic energy following Eq.~\ref{eq:dirac_radial_low_energy} in the low energy limit. For the Dirac solution, the fast core-shell electrons are accompanied by the increasing contribution of the small component shown in ~\ref{fig:orbital_speed}(b). Here we define the contribution of the small component $\zeta_{n\kappa}$ for a given orbital $n\kappa$ as:
\begin{equation}
  \zeta_{n\kappa} = \frac{\int_0^{\infty} \left(Q_{n \kappa}^2(r)\right) \mathrm{d} r}{\int_0^{\infty} \left(P_{n \kappa}^2(r)+Q_{n \kappa}^2(r)\right) \mathrm{d} r} = \int_0^{\infty} \left(Q_{n \kappa}^2(r)\right) \mathrm{d} r,
\end{equation}
since the wavefunctions are already normalized in Eq.~\ref{eq:dirac_normalized}. Fig.~\ref{fig:orbital_speed}(c-e) shows the radial charge density for three typical elements ranging from light to heavy (Si, Ag, and Au). As expected, the large component always dominates the charge distribution. In contrast, the small component is negligible for Si, but becomes noticeable for Ag and significant for Au, mainly for core-shell electrons close to the nucleus as indicated by the orbital decomposition shaded in different colours. The direct consequence of the Dirac solution for GOS is reflected in the radial integral of Eq.~\ref{eq:radial_integral_relativistic}, where the small component contributes to the transition matrix element. Further analysis of the \ce{L2}/\ce{L3} edges shows that the contribution of small components to the GOS is negligible for Si, approximately 1\% for Ag, and 3-5\% for Au (depending on the spin), which is proportional to the contribution of the small components to the charge distribution (Si: 0.03\%, Ag 0.6\%, Au 2\%).

To distinguish the contributions of the small component and relativistic kinematics, Fig. S5 in the supplementary materials presents the corresponding ratio plots as a function of atomic number for the 1s, 2s, and 2p orbitals, which are most influenced by relativistic effects. Although these ratios follow similar trends, the gap between \ce{2p_{1/2}} and \ce{2p_{3/2}} is narrower in the small component ratios than in the relativistic kinematic corrections. Notably, the ratio of orbital binding energy to rest mass energy serves as a good indicator of the small component contribution, although it differs in scaling.

\begin{figure}[htbp]
    \centering
    \includegraphics[width= 1.0\textwidth]{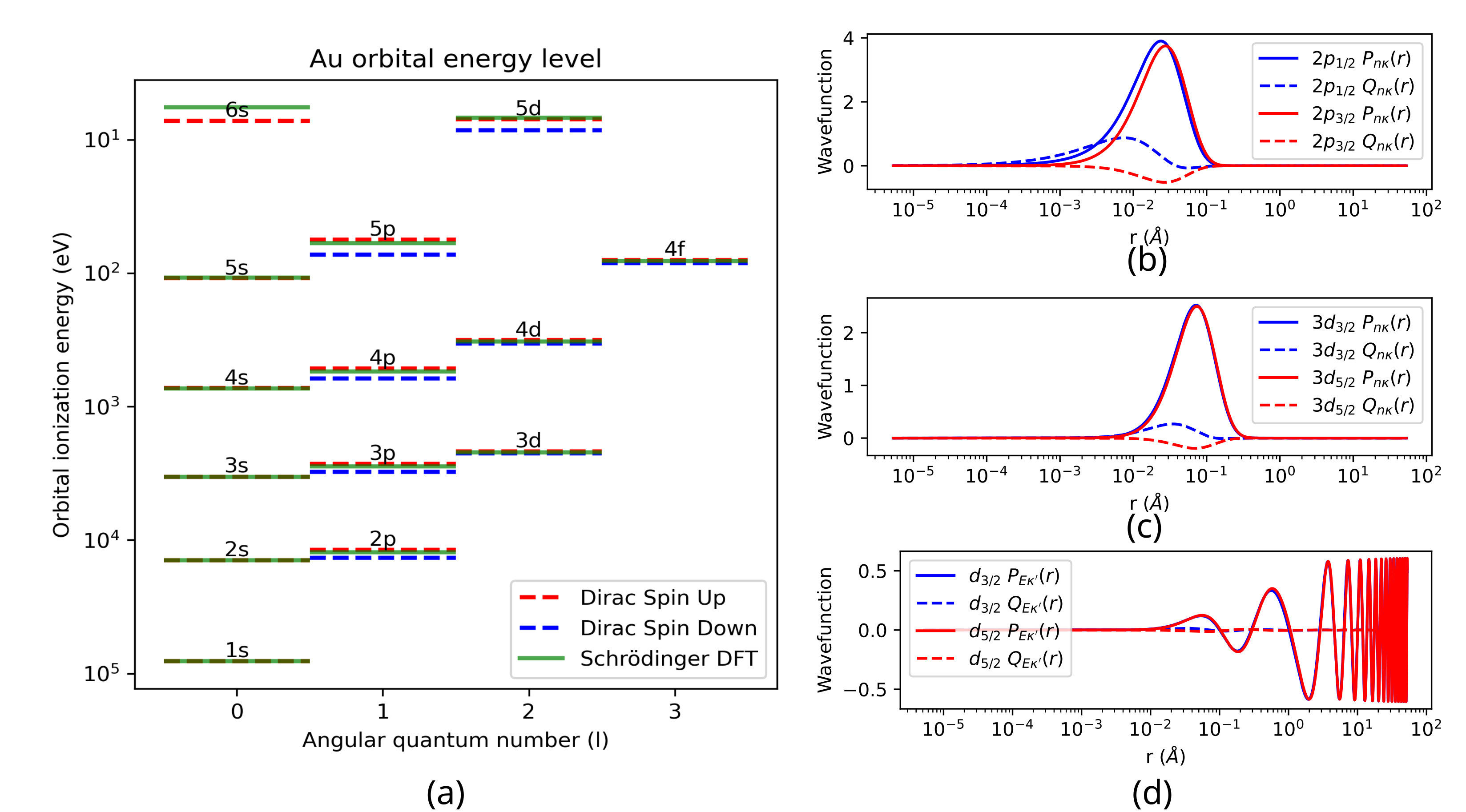}
    \caption{(a) Orbital energy level for Au as a function of angular quantum number with spin variations. The red/blue shaded line indicates spin up/down Dirac solution, while the green solid line indicates the Schrödinger all-electron DFT solution. The radial wavefunctions for (b) the 2p orbital, (c) the 3d orbital, and (d) continuum state d orbital at 10~eV with spin variations.}
    \label{fig:orbital_spin}
\end{figure} 

The spin variation also yields different wavefunctions and hence different transition matrixes and GOS, in contrast to a single matrix from the Schrödinger solution. In Fig.~\ref{fig:orbital_spin}(a), the electron ionization energy is plotted as a function of the orbital and spin variations, wherein Fig.~\ref{fig:orbital_spin}(b-d) plots the radial wavefunctions for the bound and continuum states. The significance of energy splitting and deviation from the Schrödinger solution is a direct indication of the strength of SOC, which is particularly pronounced for p orbitals and less obvious for higher orbitals (like d orbitals) in general due to electron screening (note energies plotted on a log scale). For radial wave functions of bound states, the \ce{2p_{1/2}} orbital is contracted inward compared to the \ce{2p_{3/2}} orbital for both the large and small components as shown in Fig.~\ref{fig:orbital_spin}(b). In contrast, the large components for 3d orbitals are similar regardless of the spin, while its small component is slightly contracted inward for \ce{3d_{3/2}} electron compared to \ce{3d_{5/2}} as shown in Fig.~\ref{fig:orbital_spin}(c). To calculate the matrix elements, the bound state wavefunctions will be integrated with different final continuum states for each energy, one of which is shown in Fig.~\ref{fig:orbital_spin}(d) for a final state angular quantum number $l'=2$ with spin-up/down at 10~eV above the ionization energy. The highly oscillating nature of the continuum state (with frequency increases with increasing energy loss) and spherical Bessel function (with frequency increases with increasing momentum transfer) suggests that a slight shift in the radial wavefunction of the highly localized bound state can result in a remarkable change in the integral wavefunction overlaps for bound states with different spin quantum numbers. 

\begin{figure}[htbp]
    \centering
    \includegraphics[width= 1.0\textwidth]{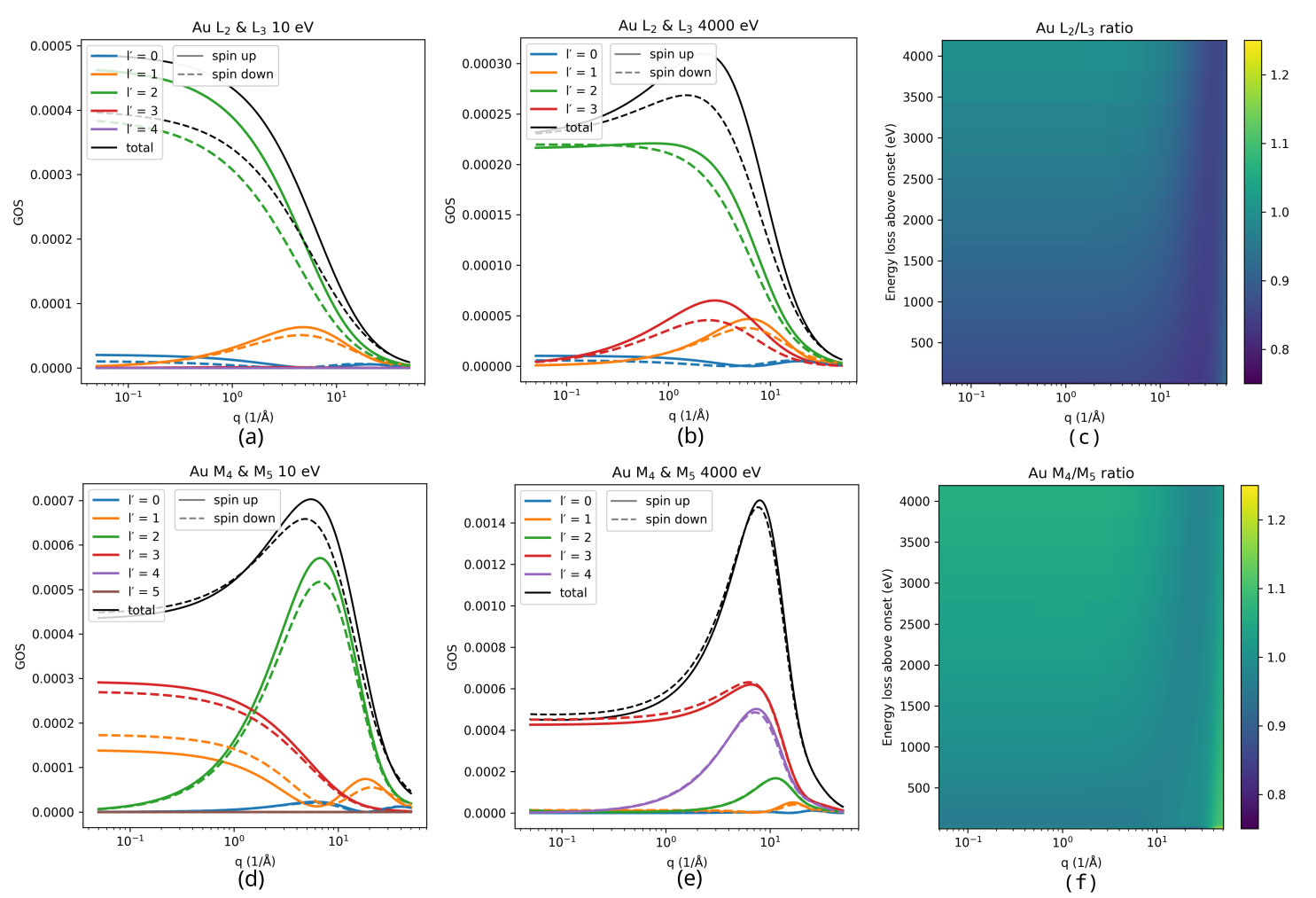}
    \caption{The decomposition of the contributions from different final states for the Au \ce{L2} and \ce{L3} case at (a) 10~eV and (b) 4000~eV above the ionization energy and similarly for the Au \ce{M4} and \ce{M5} case in (d-e). The spin-down/up GOS ratio against the energy loss and momentum transfer for (c) \ce{L2}/\ce{L3} and (f) \ce{M4}/\ce{M5} edges.}
    \label{fig:spin_ratio}
\end{figure}

To investigate how spin variation affects the GOS, Fig.~\ref{fig:spin_ratio} demonstrates the contributions of final states with different angular quantum numbers and spin variations, plotted as a function of energy loss and momentum transfer. Fig.~\ref{fig:spin_ratio}(a-b) shows the decomposition of the contributions from final states for the Au \ce{L2}/\ce{L3} case at 10~eV and 4000~eV above the edge onset. At low energy loss, the dipole transition $\Delta l' = |l-l'| = 1$ dominates the GOS as expected. However, for high energy loss, there are notable contributions from final states with other $l'$ leading to the emergence of a peak at high momentum transfer known as the Bethe ridge \cite{egerton2011electron}. In the case of Au, the difference of spin reaches 20\% for \ce{L2}/\ce{L3} edges, but it can be higher for a heavier element (like actinium can be over 40\%). This difference in GOS is not surprising, as the spin-down electron is closer to the nucleus with a higher ionization energy due to SOC, its excitation is expected to have a lower transition probability compared to the spin-up electron. In contrast, the ratio of spin-down/up is close to 0.95 for \ce{M4}/\ce{M5} edges which is a result of negligible spin variations in their wavefunctions shown in Fig.~\ref{fig:orbital_spin}(c). 

%

\subsection{Relativistic electrodynamics for EELS cross-section calculations}
\label{sec:relativistic_electrodynamics}
\begin{figure}[htbp]
    \centering
    \includegraphics[width= 0.8\textwidth]{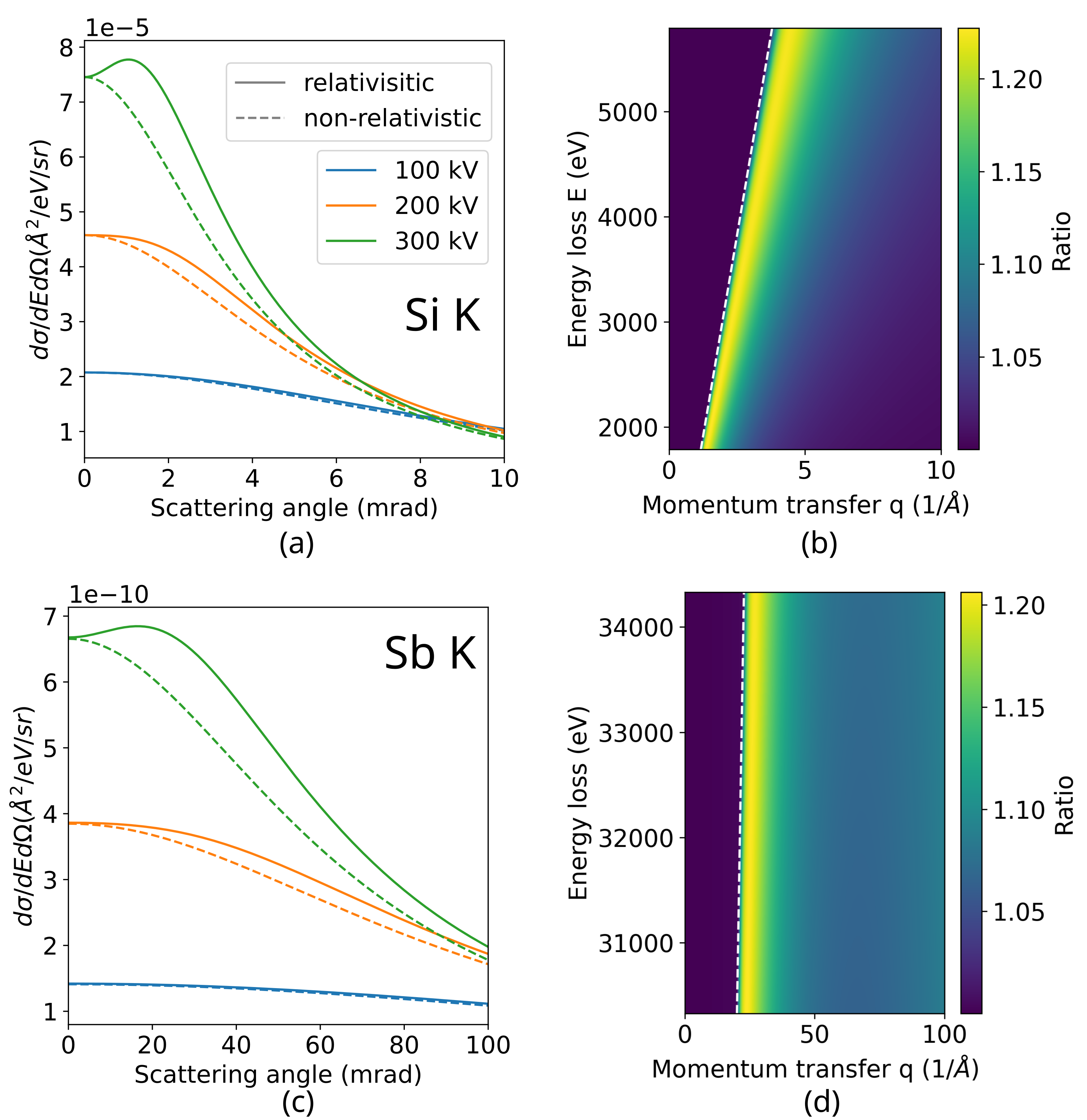}
    \caption{Relativistic and non-relativistic double differential cross sections for (a) silicon and (c) antimony K-ionization at 10~eV above the ionization energy with the incident electron of 100 - 300~keV. The ratio of relativistic to non-relativistic differential double differential cross sections for (b) silicon and (d) antimony K-ionization at 300~keV incident beam energy as a function of momentum transfer and energy loss. The white dashed line indicates the minimum momentum transfer.}
    \label{fig:relativistic_electrodynamics_correction}
\end{figure} 

\begin{figure}[htbp]
    \centering
    \includegraphics[width= 0.8\textwidth]{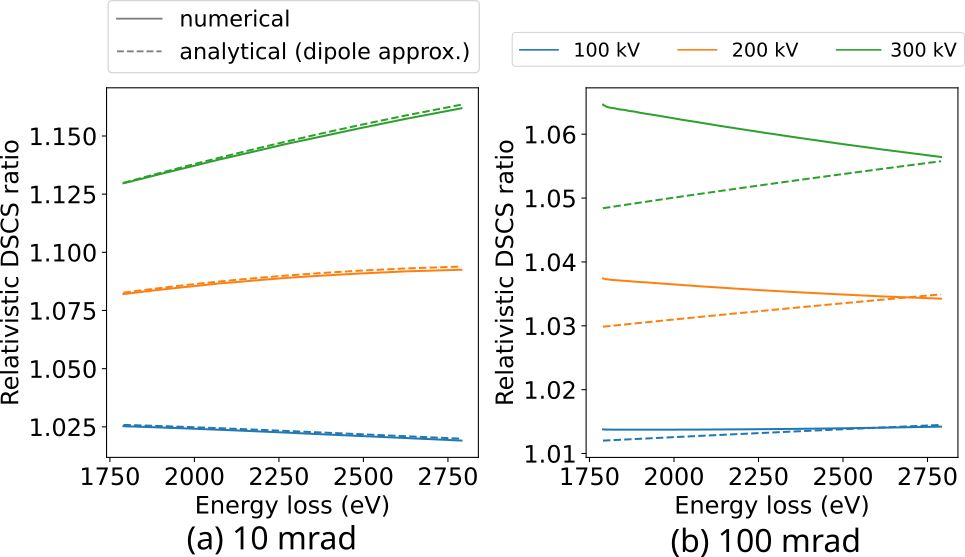}
    \caption{The relativistic differential scattering cross-section (DSCS) correction ratio of the Si K edge against the energy loss at different acceleration voltages (100-300~keV) with (a) 10~mrad and (b) 100~mrad collection angles using the full numerical solution and the analytical solution with dipole approximation.}
    \label{fig:correction_intergrate_collection_angle}
\end{figure}

The relativistic effects in electrodynamics significantly enhance the scattering cross-section, particularly noticeable at low angles and high energies. Fig.~\ref{fig:relativistic_electrodynamics_correction}(a) shows the relativistic and non-relativistic double differential cross sections for silicon K-edge at different acceleration voltages, showing a small angle peak, which becomes more pronounced with increasing acceleration voltage, negligible at 100~keV but not for higher voltages. As explained in previous studies \cite{knippelmeyerRelativisticIonisationCross1997,Schattschneider2005a,Dwyer2006}, this relativistic correction is obtained when applying Lorentz gauge for a moving charge in Eq.~\ref{eq:DDSCS_relativistic}-\ref{eq:DDSCS_relativistic_simplified}. It also leads to the interference of the longitudinal and transverse components of the excitation, resulting in anisotropic differential cross-sections which is critical when measuring near-edge fine structures \cite{Schattschneider2005a}. Fig.~\ref{fig:relativistic_electrodynamics_correction}(b) shows the ratio of relativistic to non-relativistic differential cross sections for silicon K-ionization at 300~keV incident beam energy as a function of momentum transfer and energy loss, in contrast to the single energy loss in Fig.~\ref{fig:relativistic_electrodynamics_correction}(a), which agrees with previous results \cite{knippelmeyerRelativisticIonisationCross1997,Dwyer2006}. The correction is significant at the small scattering angles and varies slowly against increasing energy loss. A white dashed line in Fig.~\ref{fig:relativistic_electrodynamics_correction}(b) indicates the minimum momentum transfer. Similar results are observed for the antimony K-edge in Fig.~\ref{fig:relativistic_electrodynamics_correction}(c-d), displaying a comparable pattern; however, the relativistic effects extend to what is conventionally considered a high collection angle (i.e., 100 mrad). Further investigation reveals that the relationship between the DDSCS and scattering angle is well characterized when expressed in terms of the characteristic scattering angle, which scales with energy loss. To appreciate this effect on the spectra under conventional experimental conditions, we can integrate DDSCS the scattering within the EELS collection aperture via Eq.~\ref{eq:diff_cross_section_integral_q} to differential scattering cross-section. To check we have the correct implementation of the relativistic correction, Fig.~\ref{fig:correction_intergrate_collection_angle} plots the relativistic correction ratio of the Si K-edge at different acceleration voltages with (a) 10~mrad and (b) 100~mrad collection angles using numerical calculation in Eq.~\ref{eq:DDSCS_relativistic_simplified} and analytical expression with dipole approximation in Eq.~\ref{eq:relativistic_dipole_correction}. The numerical calculation of the relativistic correction ratio is consistent with the analytical dipole solution in Eq.~\ref{eq:relativistic_dipole_correction} for small angles of 10~mrad, yielding up to a 15\% increase in the cross-sections at 300~keV. However, with increasing collection angle, the relativistic double differential cross-sections are very close to the non-relativistic ones (i.e. only about 5\% for 300~keV and 100~mrad) as shown in Fig.~\ref{fig:relativistic_electrodynamics_correction}. It is important to note that the dipole approximation is expected to fail at high angles and one has to use numerical relativistic corrections. Overall, we want to emphasize that the relativistic correction is important for all elements (in contrast to the relativistic atomic orbital effects, which are only obvious for core-shell electrons of heavy elements). Since this relativistic effect varies with energy loss and momentum transfer, the elemental quantification will be affected differently for different edges under varying experimental conditions. These corrections should be routinely performed when quantifying EELS with typical TEM acceleration voltages at 200-300~keV, particularly when using a small collection aperture.

\subsection{Comparison of the Dirac-based GOS database with others}
\label{sec:compare_gos}
\begin{figure}[htbp]
    \centering
    \includegraphics[width= 1.0\textwidth]{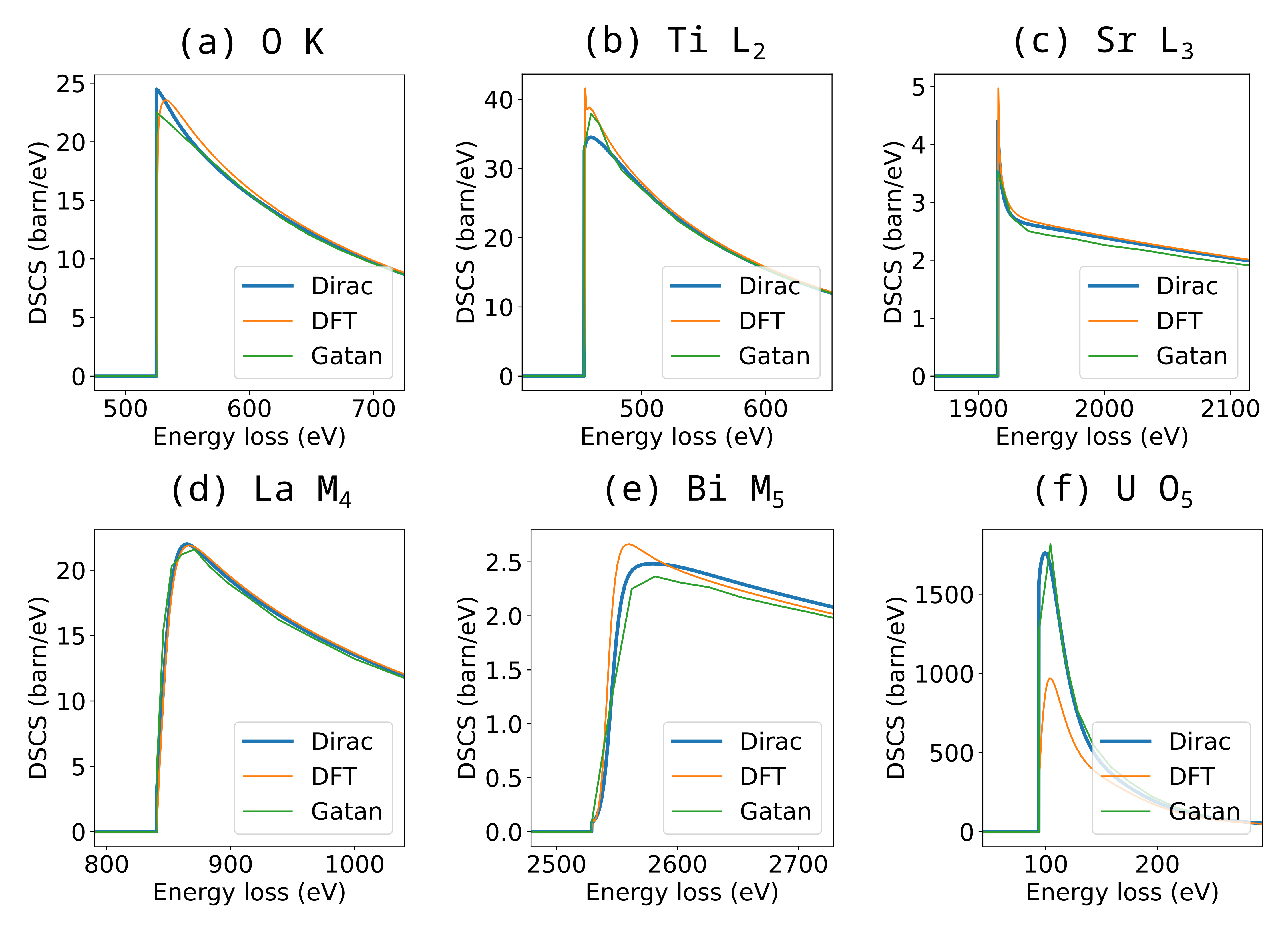}
    \caption{Comparison of the GOS databases for (a) O K, (b) Ti \ce{L2}, (c) Sr \ce{L3}, (d) La \ce{M4}, (e) Bi \ce{M5} and (f) U \ce{O5} edges using different GOS databases with 300~keV incident beam, parallel illumination, and 50~mrad collection angle.}
    \label{fig:compare_gos}
\end{figure} 

We compared our Dirac-based GOS with the existing GOS database used in the EELS community, including the Gatan Dirac solutions \cite{Rez2002,Rez2003}, and DFT solutions based on the Schrödinger equation \cite{Segger2019,Segger2023}. For a fair comparison evaluation of the GOS databases themselves, the relativistic electrodynamics effect in Sec.~\ref{sec:relativistic_electrodynamics} is not included in this comparison. In Fig.~\ref{fig:compare_gos}, the calculated spectra for O K, Ti \ce{L2}, Sr \ce{L3}, La \ce{M4}, Bi \ce{M5}, and U \ce{O5} edges using different GOS databases with 300~keV incident beam energy, parallel illumination, and a 50~mrad collection angle. Overall, these spectra exhibit similar shapes and absolute amplitude except for a few details. Specifically, the Gatan spectra are more fragmented due to the limited sampling in both energy and momentum, as shown in Fig.~\ref{fig:gos_sampling}. The Schrödinger cross-sections are in good agreement with the Dirac ones for light elements where SOC is expected to be weak. However, notable differences are observed in the heavy elements for the  Bi \ce{M5} and U \ce{O5} edges. For the U \ce{O5} edge, Gatan and Dirac curves are in agreement while the DFT curve deviates away at the near edge region. Also, we could see a sharp peak at the start of the DFT Ti \ce{L2} edge. Since those GOS are computed using different internal numerical routines, it is difficult at this stage to pinpoint whether the discrepancy is attributed to numerical issues or the underlying physics. It could be useful to calculate the Schrödinger and Dirac solutions and construct the GOS under the same protocols to trace the causes. We noticed that there are packages (e.g. RIDIAL\cite{Salvat2019} and dftatom \cite{vcertik2013dftatom}) that can compute both the Schrödinger and Dirac atomic radial wavefunctions, which will enable the direct comparisons. 

We expect that the inner core-shell edges (i.e. K and L-edge) of heavy elements will offer a better opportunity to observe the relativistic effects, although this corresponds to extremely high energy loss that is typically not explored often in EELS. Extremely high loss EELS is mainly limited by the low signal-to-noise from the small cross-sections of those core electron edges of heavy elements and the loss in energy resolution due to the geometric and chromatic aberrations. Recent hardware developments in EELS aberration correction and detectors enable extremely high energy losses with examples including the W L3 ($\approx$ 10 keV) \cite{maclarenEELSVeryHigh2018} and Sb K ($\approx$ 30 keV) \cite{lazarEELSVeryHigh2023a} edges and comparable with XAS measurement. With the development of direct electron cameras and advanced spectrometers, experimental verification of the Dirac cross-sections can be expected in the future.


\subsection{Use of the Dirac-based GOS database}
\label{sec:how_to_use}
The Dirac-based GOS database is made available open-source on Zenodo under a CC-BY license \cite{Zhang2023}. Similar to using all GOS databases, quantifying the EELS cross-sections involves a three-step process. 

Step 1, we need to compute the DDSCS from the GOS database given in Eq.~\ref{eq:ddscs} for the non-relativistic scattering but with atomic orbitals obtained from the Dirac equation. Then we use Eq.~\ref{eq:DDSCS_relativistic_simplified} to account for the relativistic nature of the fast incoming electron. Step 2, we need to integrate the DDSCS within the EELS collection aperture to obtain the differential cross-sections in Eq.~\ref{eq:diff_cross_section_integral_q} for parallel illumination. If convergent beam illumination is used in STEM-EELS, we have to correct it with a geometric cross-correlation function in Eq.~\ref{eq:diff_cross_section_integral_q_BF}  \cite{kohl1985simple}. The momentum space integration for both parallel and convergent beam illumination is implemented in pyEELSModel. Step 3, the computed differential cross-sections for all edges are fitted together to the experimental spectrum for quantification. During the model-based fitting process, the core-loss spectrum can be convoluted with the low-loss spectrum to consider the source energy dispersion and plural plasmon scattering. The fitting procedure can be performed using the existing packages such as EELSModel, HyperSpy, or Gatan Digital Micrograph, which now all support the Dirac-based GOS. We also provide a Jupyter notebook \cite{Zhang_gos_demo} to demonstrate the fitting process of the Dirac-based GOS for the Ti \ce{L2}/\ce{L3} and O K edges in \ce{SrTiO3} using pyEELSModel \cite{daen_jannis_2024_10992986}. We encourage inquiries from the community and industry to integrate the Dirac-based GOS into their software packages.

The GOS developed in this work has already been applied in several recent studies, including low-voltage STEM-EELS \cite{dumaresqElementalQuantificationUsing2024}, deep learning for core-loss edge identification \cite{annysDeepLearningAutomated2023b}, and fine-structure quantification with improved precision constrained by the Bethe sum rule \cite{jannisImprovedPrecisionAccuracy2024}.

\section{Conclusions and outlook}
\label{sec:conclusions}
EELS is a powerful technique for the characterization of materials composition and electronic structures. The accuracy of the EELS quantification relies on the precise calculation of the cross-sections facilitated by the GOS database, which computes the transition probability as a function of energy loss and momentum transfer. In this study, we developed an open-source Dirac-based GOS database for all elements and shells for large energy and momentum space with fine sampling. Unlike previous GOS databases relying on the Schrödinger equation, the Dirac-based GOS database better captures the relativistic effects. We demonstrated that the ionization energies calculated from the Dirac equation align well with the experimental measurements with a relative error of $0.2\%\pm 7\%$, including the energy splitting caused by SOC. Notably, the large component of the Dirac radial wave is shown to be equivalent to the non-relativistic Schrödinger solution at low energies, while the small component of the Dirac solution becomes significant for the core-shell electrons of heavy elements. We also showed that SOC leads to different wavefunctions and hence different transition matrix elements and GOS for spin variations. We compared the Dirac-based GOS database with existing GOS databases commonly used in the EELS community. Despite fine differences usually at the near-edge region, these spectra are generally similar for the K and L edges of light elements or the higher edges of heavy elements. But for the core-shell electrons of heavy elements, theoretical results suggest considerable difference for spin up/down (up to 20\% for Au and 40\% for Ac), which is rarely measured experimentally by EELS. Further experimental work is needed to verify the Dirac-based GOS for absolute cross-sections of the heavy elements at high energy loss, which requires an advanced EELS spectrometer and detector capable of this precise measurement. Overall, the Dirac-based GOS database is expected to be useful for microanalysis with relativistic effects considered across the periodic table. In addition, the differential cross-sections could be also useful for the Monte Carlo simulations of electron scattering.

The calculations in this work are dedicated to the GOS database for EELS microanalysis. However, this computational framework can be easily extended to generate the oscillator strength database for XAS. As the momentum transfer is negligible for photons, the optical dipole oscillator strength is not dependent on the scattering angle. This can be obtained by substitution of the Hamiltonian in the transition matrix of Fermi's golden rule, which is the Coulombic interaction for an electron, with the position operator and the polarization vector of the X-ray beam. Indeed, EELS and XAS are often used in tandem to provide complementary information on materials. A pair of EELS GOS and XAS oscillator strength databases should be of interest to material scientists who use both techniques. 

In analytical electron microscopy for microanalysis, the most common techniques are energy-dispersive X-ray spectroscopy (EDS) and EELS. The EDS collects the characteristic X-ray generated during the de-excitation of the core-shell electron, which is a physically coupled process and therefore its theoretical cross-sections are deeply related to the EELS calculations presented here. Experimentally, EDS elemental mapping is much simpler than EELS and favoured by material scientists. However, accurate and precise EDS quantification is not easy, particularly at the atomic scale, due to various theoretical and experimental challenges. We have generated the Dirac-based ionization potential database for atomic EDS calculations in the Bloch wave or multislice algorithms, which includes the relativistic effects from both the orbital and fast incident electron. The Dirac-based EDS database will be made available to the community in the future. 

We notice that the ionization energy depends on the chemical environment, which is also expected to change the cross-sections as well. Thus, the calculation of the GOS for different oxidation states is currently under development. Moreover, our current analysis, limited to continuum states of individual atoms, fails to encapsulate the intricate unoccupied DOS inherent in complex, multi-atom systems, thereby neglecting the fine structures in the near-edge region. Given that the near-edge region often contains most of the signal counts against background and noise, the exclusion of physics-based fine structure from spectrum fitting represents a significant loss of information. Previous studies have shown that the white line caused by transitions to discrete bound states can contribute significantly to the cross-sections \cite{rezContribution1992}. Hence, we are strongly motivated to incorporate the unoccupied bounded states in our future generation of theoretical EELS cross-sections.

\section{Data availability}
The database is available on Zenodo under a CC-BY license \cite{Zhang2023}.

\section*{Acknowledgement}
The authors acknowledge financial support from the Research Foundation Flanders (FWO, Belgium) through Project No.G.0502.18N. This project has received funding from the European Research Council (ERC) under the European Union’s Horizon 2020 research and innovation programme (Grant Agreement No. 770887 PICOMETRICS and No. 823717 ESTEEM3). The authors acknowledge the computational resources from Vlaams Supercomputer Centrum (VSC) through Tier-1 project No.2023.023. ZZ acknowledges the travel funding from FWO (Grant No.V463823N) and ENEN2plus (Grant No.0000001160) and the consultation with Dirk Van Dyck, Peter Schattschneider, Christian Dwyer, and Christian Rossouw. Special thanks are extended to Francesc Salvat for generously sharing the GOSAT code and data for results comparison. ZZ thanks Linfeng Zhang and Chu-ping Yu for reading the manuscript and offering helpful comments.

\bibliography{references}
\end{document}